\documentclass{aa}
\usepackage[varg]{txfonts}

\usepackage{natbib}
\bibpunct{(}{)}{;}{a}{}{,}

\usepackage{multirow}

\begin{document}

\title{Resolving white dwarf binaries within globular clusters with LISA}

\author{Wouter G. J. van Zeist\inst{\ref{radb},\ref{leid},\ref{auck}}\thanks{Corresponding author: wouter.vanzeist@astro.ru.nl}
\and Gijs Nelemans\inst{\ref{radb},\ref{sron},\ref{leuv}}
\and Shu-Xu Yi\inst{\ref{ihep},\ref{ucas}}
\and Simon F. Portegies Zwart\inst{\ref{leid}}
}

\institute{Department of Astrophysics/IMAPP, Radboud University, PO Box 9010, 6500 GL, Nijmegen, The Netherlands\label{radb}
\and Leiden Observatory, Leiden University, Einsteinweg 55, 2333 CC, Leiden, The Netherlands\label{leid}
\and Department of Physics, University of Auckland, Private Bag 92019, Auckland, New Zealand\label{auck}
\and SRON, Netherlands Institute for Space Research, Niels Bohrweg 4, 2333 CA, Leiden, The Netherlands\label{sron}
\and Institute of Astronomy, KU Leuven, Celestijnenlaan 200D, B-3001, Leuven, Belgium\label{leuv}
\and Key Laboratory of Particle Astrophysics, Institute of High Energy Physics, Chinese Academy of Sciences, Beijing 100049, China\label{ihep}
\and University of Chinese Academy of Sciences, Chinese Academy of Sciences, Beijing 100049, China\label{ucas}
}

\date{Received XXX / Accepted YYY}

\abstract{Globular clusters (GCs) around the Milky Way (MW) are expected to host white dwarf (WD) binaries emitting gravitational waves that could be detectable by LISA.}
{Our aim is to investigate whether LISA can resolve WD binaries in GCs well enough in terms of sky location and distance that they can be distinguished from binaries in the MW disc.}
{We used a sample of 20 of the most massive GCs around the MW and simulated LISA's sky location and distance measurement errors for WD binaries in these GCs using the software package \textsc{GWToolbox}. We did this in the context of a model of the LISA-detectable binaries in the MW from the population synthesis code \textsc{SeBa}.}
{We find that for five of the GCs in our sample, binaries in the GC could be easily distinguished from MW disc binaries using the sky location alone; for another five, binaries in the GCs could be distinguished using a combination of LISA's sky location and distance measurements; and for the final ten, binaries in the GCs could not be distinguished from overlapping MW disc binaries. The results depend strongly on the sky locations of the GCs, with GCs far away from the Galactic plane being easy to resolve, while GCs close to the Galactic centre overlap with many MW disc binaries. The most promising GC for finding a WD binary that could be resolved to that GC, based on sky location and GC mass, is 47 Tucanae.}
{}


\titlerunning{Resolving white dwarf binaries within globular clusters with LISA}
\authorrunning{W. G. J. van Zeist et al.}

\maketitle

\section{Introduction} \label{intro_locpaper}

The Laser Interferometer Space Antenna \citep[LISA;][]{lisa_l3,lisa_redbook} is a space-based gravitational wave (GW) observatory that is currently under development for launch in 2035. It will be able to detect GWs at frequencies from approx. 0.1 mHz to 0.1 Hz, a lower frequency band than the currently operational, ground-based GW detectors of the LIGO–Virgo–KAGRA (LVK) Collaboration \citep{gwtc3}.

LISA has been predicted to be able to detect many different types of GW sources \citep{lisa_astrophysics,lisa_cosmology}, one important class of which are stellar-origin compact binaries, which may contain black holes (BHs), neutron stars (NSs) or white dwarfs (WDs). There have been many studies about the populations of compact binaries that LISA may detect \citep[e.g.][]{paczynski1967,lipunov1987,hils1990,schneider2001,nelemans2001,ruiter2010,belczynski2010,nissanke2012,korol2017,lamberts2018,lamberts2019,breivik2020,eccentricity_lau,bpassmilkyway}; overviews are given in Sect. 1 of \citet{lisa_astrophysics} and Sect. 4.3 of \citet{breivik2025}. Most of the compact binaries that LISA will detect will be within the Milky Way (MW), and the most numerous type will be double white dwarf (DWD) binaries \citep{lisa_astrophysics}.

In addition to the MW disc itself, globular clusters (GCs) orbiting the MW have also been considered as potential hosts of binary GW sources for LISA \citep[e.g.][]{lisa_gc1,ivanova2006,willems2007,kremer2018_gc_lisa,wouter_clusters_clouds}. The aforementioned studies are based primarily on simulations, but there are also potentially LISA-detectable binaries in GCs known from electromagnetic (EM) observations, such as 47 Tuc X9 \citep{hertz1983,verbunt1998,grindlay2001}.

GCs are hundreds of thousands to millions of times less massive than the MW, and each occupies only a small portion of the sky seen from Earth. The stellar populations of GCs are also distinct from those in the MW disc in a number of ways: firstly, they are more homogeneous in terms of age and metallicity, though they can contain multiple populations of different ages \citep[see e.g.][]{bastian2018,milone2022}. Secondly, the stellar density in GCs is higher than in the MW disc, which increases the chance that stellar systems will have gravitational encounters with each other. These gravitational encounters can affect the properties of the stellar population, and consequently of the compact binaries formed from the stellar population, in various ways \citep[see e.g.][]{heggie1975,heggie2003,ivanova2006,benacquista2013,kremer2020}.

Various studies have made predictions of how many binaries LISA will be able to detect in GCs, with their results differing depending on the models used for simulating stellar evolution and GC evolution. \citet{willems2007}, using the GC evolution code \textsc{cmc} \citep{joshi2000,rodriguez2022} and the stellar evolution code \textsc{startrack} \citep{startrack_1,startrack_2}, predicted several dozen LISA-detectable DWDs across all MW GCs combined. \citet{kremer2018_gc_lisa}, using \textsc{cmc} and the stellar evolution codes \textsc{sse} \citep{hurley2000} and \textsc{bse} \citep{hurley2002}, predicted 14 to 21 LISA-detectable binaries across all MW GCs combined, including 4 to 6 DWDs. \citet{hellstrom2025}, using \textsc{bse} and the GC simulation code \textsc{mocca} \citep{hypki2013,hypki2025}, predicted 5 to 15 LISA-resolvable DWDs across all MW GCs combined.

By contrast, \citet{wouter_clusters_clouds}, using the stellar evolution code \textsc{bpass} \citep{bpass1,bpass2}, predicted that the total number of LISA-detectable binaries across all MW GCs would be 0 or 1. However, it should be noted that \textsc{bpass} generally underpredicts the number of DWDs in the LISA band by at least an order of magnitude when compared to observations from EM surveys \citep{wouter_em_comparison}.

Aside from the number of LISA-detectable binaries in GCs, another question is whether LISA, if it detects a binary in a GC, would be able to identify that binary as being part of that GC. This question arises because numerous simulations predict that LISA will detect ten thousand or more binaries, mostly DWDs, in the MW disc \citep[e.g.][]{ruiter2010,korol2017,lamberts2019,delfavero2025}, and each of these binaries will be detected with a finite resolution in sky location and distance. Therefore, the uncertainty range in the measured sky locations and distances of the binaries in the MW disc could overlap the GC, which could make it difficult to distinguish binaries in GCs from those in the MW disc; though, we note that there are also other binary parameters that may be usable to distinguish GC and MW disc DWDs, such as eccentricity \citep{hellstrom2025}, which we discuss more in Sect. \ref{discussion_eccentricity}.

Our goals in this study are to investigate LISA's sky localisation of WD binaries, and to judge how often these overlaps may occur and what their impact would be in terms of distinguishing GC and MW disc binaries, in order to gain insight on the question: how well could LISA identify binaries in GCs as belonging to those GCs? We note that \citet{xuan2025} have recently analysed LISA's sky localisation of BH binaries in GCs, and found that LISA could indeed localise these binaries to their GC if their signal-to-noise ratio (S/N) > 20. However, LISA will detect orders of magnitude more WD binaries than BH binaries in the MW disc \citep{lisa_astrophysics} and, as previously stated, there are also expected to be LISA-detectable WD binaries in GCs – potentially as many as, if not more than, BH binaries \citep{lisa_gc1,kremer2018_gc_lisa,wouter_clusters_clouds}.

\section{Method}

\subsection{Sample of globular clusters}

For our investigation of LISA's sky localisation of binaries in GCs, we used a sample of 20 GCs. Specifically, we used the sample of GCs previously compiled in \citet{wouter_gw_spectral}, containing the 20 most massive GCs within 10 kpc of Earth from the catalogues of \citet{catalogue_baumgardt1} and \citet{catalogue_baumgardt2}. We list the coordinates and distances of each of these GCs in Table \ref{param_table}; other information on these GCs can be found in Table C2 of \citet{wouter_gw_spectral}.

\subsection{Milky Way model}

To investigate how well LISA can distinguish binaries in GCs from those in the MW disc, we needed a model of the LISA sources in the MW disc. For this, we used the galaxy model of \citet{korol2017}. This galaxy model consists of \textsc{SeBa} stellar evolution models \citep{portegies1996,nelemans_seba,seba_toonen} combined with models of the MW's star formation history and structure from \citet{nelemans_seba,nelemans2004} and \citet{toonen2013}, which were based on the galaxy evolution model of \citet{boissier1999}. \citet{korol2017} produced two versions of their galaxy model with different assumptions about common-envelope evolution, named ``$\alpha\alpha$'' and ``$\gamma\alpha$'' respectively, of which we used the latter. The \textsc{SeBa} models use an initial mass function from \citet{kroupa1993} and initial binary parameters from \citet{heggie1975} and \citet{abt1983}. The motivation for using this particular galaxy model is that \citet{wouter_em_comparison} showed that its predictions are largely consistent with existing electromagnetic observations of DWDs.

The galaxy model contains a total of 26,433,152 DWDs with a minimum GW frequency (twice the orbital frequency) of 10$^{-4}$ Hz. However, many of these DWDs will not be individually resolvable by LISA; those DWDs would still contribute to the foreground confusion noise in the detector, but are not relevant for this study, as we are only interested in those binaries that LISA can localise on the sky.

To select only those DWDs that are resolvable, we used the software package \textsc{GWToolbox} \citep{gwtoolbox1,gwtoolbox2,gwtoolbox3} to calculate the signal-to-noise ratio (S/N) of each of the DWDs. We then applied a simple, loose threshold to select the detectable binaries: a S/N > 5, calculated with the inclination and polarisation angles set to zero. The reason we used a loose threshold is that DWDs with a low S/N will be smeared out in terms of sky location and distance uncertainty, and therefore these borderline systems will not be important for our analysis of how well GC DWDs can be distinguished from those in the MW disc.

After applying the S/N threshold, we obtained a dataset of 26,563 potentially resolvable DWDs. This is similar to the figure of 25,000 LISA-detectable DWDs that \citet{korol2017} themselves found, using a different code to calculate the S/Ns.

\subsection{Computing sky location and distance errors} \label{method_compute_errors}

For each of the 26,563 resolvable systems from the \textsc{SeBa} galaxy model, we used \textsc{GWToolbox} to calculate their sky location and distance errors; that is, if LISA observed these systems, the uncertainties that LISA's measurements of their sky locations and distances would have. \textsc{GWToolbox} calculates these parameter uncertainties using Fisher information matrices \citep{gwtoolbox1}, based on the methods of \citet{shah2012,shah2014}.

When calculating the uncertainties, we set the inclination angle of each binary as 0.6$\pi$ rad. We chose this value after comparing the sky location and distance errors output by \textsc{GWToolbox} to the data in Table 3.1 of \citet{lisa_redbook}. The aforementioned table contains the results of LISA parameter estimation calculations for three example systems (one WD–WD binary, one NS–NS and one BH–BH) as calculated by the LISA Figures of Merit effort.

The comparison is illustrated in the appendix in Fig. \ref{plot_redbook}. For the WD and NS binaries, the sky location and distance errors from \textsc{GWToolbox} (using an inclination of 0.6$\pi$ rad) are consistent with the \citet{lisa_redbook} values, except at a frequency of 1 mHz. However, the S/Ns of the example WD and NS binaries at 1 mHz are 0.16 and 1.53 respectively, and so they are not comparable to the \textsc{SeBa} systems we are analysing that have S/N > 5. The sky location errors of the example BH binary are also not completely consistent between \textsc{GWToolbox} and \citet{lisa_redbook}, but these BH binaries have far higher S/N and tighter localisation than the WD binaries we are considering (their sky errors are far smaller than 1 deg$^2$, while our results below in Table \ref{overlap_table} will show that only a small fraction of the \textsc{SeBa} binaries are below this threshold).

\subsection{Test sources in GCs} \label{method_test_sources}

\begin{figure*}
    \centering
    \includegraphics[width=0.33\linewidth]{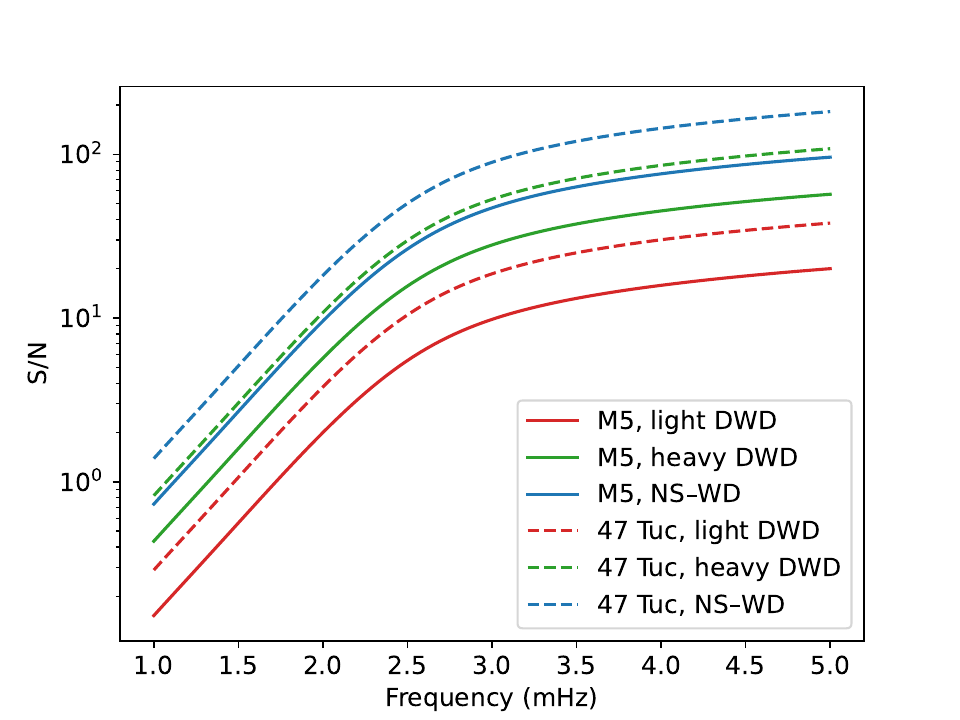}
    \includegraphics[width=0.33\linewidth]{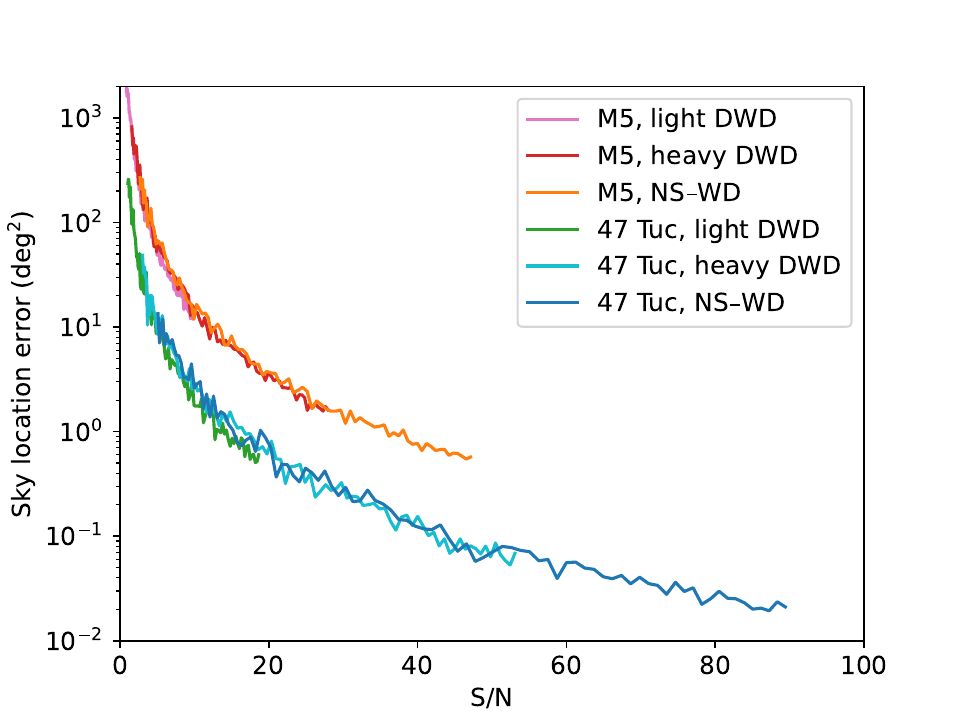}
    \includegraphics[width=0.33\linewidth]{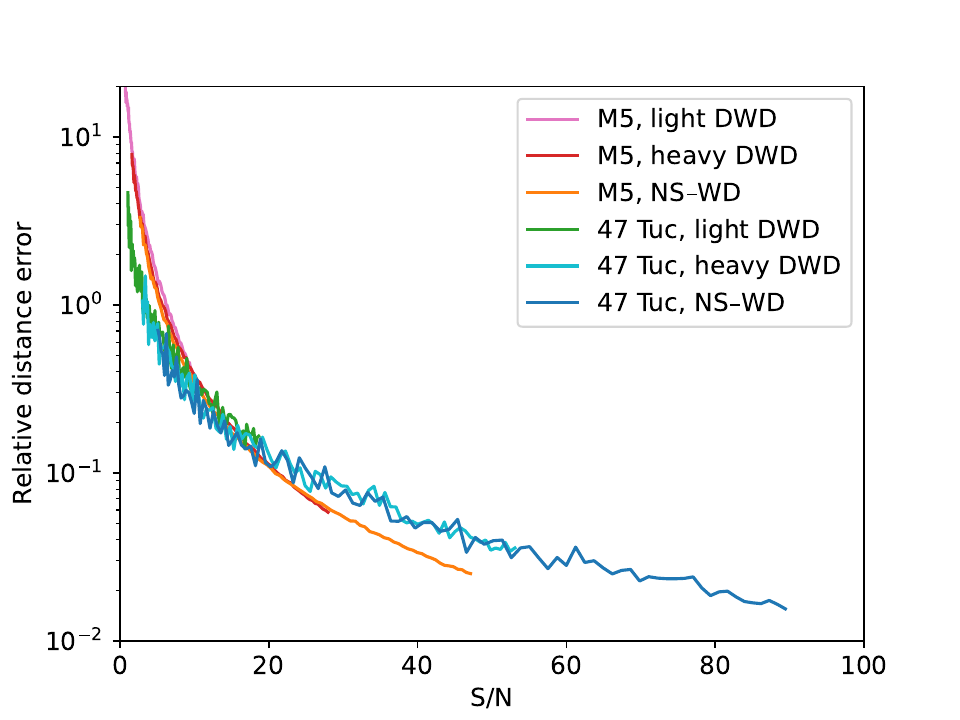}
    \caption{Illustrations of the relationship between the properties of a binary and its LISA measurement uncertainties, for three different test sources in two different GCs. The first panel shows how a binary's S/N varies with its frequency. The second shows how sky location error varies with S/N, and the third how relative distance error varies with S/N.}
    \label{plot_test_sources}
\end{figure*}

We have a model of the LISA-detectable DWDs in the MW disc, but to evaluate whether WD binaries in GCs can be distinguished from these, we also needed to model the GC binaries. To do this, we computed the uncertainties for several test sources placed at the sky locations and distances of our GCs.

As mentioned in Sect. \ref{intro_locpaper}, there is a large variation between the predictions of the number of LISA-detectable DWDs in GCs from different models. Therefore, we did not make any specific assumptions about the number of sources in a GC and simply placed a single test binary in a GC at a time. We used three types of test binaries: light DWDs (0.4 and 0.3 M$_{\odot}$), heavy DWDs (0.7 and 0.6 M$_{\odot}$) and NS–WDs (1.35 and 0.6 M$_{\odot}$). As in Sect. \ref{method_compute_errors}, we assume the binaries have an inclination of 0.6$\pi$ rad.

As an aside, we note that simulations of GC dynamics, such as \citet{kremer2021}, have suggested that dynamically formed DWDs in GCs tend to have heavier component masses than those in the field, with a mass distribution extending to greater than 1.0 M$_{\odot}$. The DWDs at the high end of this distribution would therefore be more massive than our ``heavy DWD'' test binary, but our ``NS–WD'' test binary can be used as a proxy for such systems, as the masses are similar and the parameter estimation calculations in \textsc{GWToolbox} do not depend on the type of objects in the binary.

Fig. \ref{plot_test_sources} shows how LISA's measurements of our test binaries vary with the properties of the binary and the GC. The first panel shows how S/N varies with GW frequency (from 1 to 5 mHz) for the three test binaries in two different GCs from our sample: M5 (with an equatorial sky location and a distance of 7.48 kpc) and 47 Tuc (with a polar sky location and a distance of 4.52 kpc). We can see that a) S/N increases with frequency within this frequency range, b) at the same frequency, heavier binaries have a higher S/N, c) at the same frequency, the S/Ns are higher for binaries in 47 Tuc than in M5 due to the GCs' different distance and sky location.

In the second panel, we still varied the frequency of the binaries between 1 and 5 mHz, but instead plotted how the size of the sky location uncertainty area varies with S/N. Similarly, in the third panel, we plotted how the relative distance error (the absolute distance error divided by the actual distance to the source) varies with the source. We can see that the uncertainties decrease in size as S/N increases, as expected. However, the relationship between S/N and the uncertainties is not the same between the different GCs. Another observation is that within each GC, the same S/N gives the same distance and sky location error regardless of the type of the binary. This is relevant as it means that, in the rest of this paper, when we talk about the sky location or distance errors of a test binary with a given S/N, we do not need to specify whether it is a light DWD, heavy DWD or NS–WD. We note that the lines in the second and third panels are somewhat jagged due to the finite resolution of the Fisher analysis used by \textsc{GWToolbox} to calculate the measurement uncertainties.

\subsection{Overlaps in sky location error} \label{method_overlap_loc}

For each of the 20 GCs in our sample, we wanted to find the DWDs in the MW disc model that ``overlap'' with the location of the GC. To do this, for each of the 26,563 DWDs from the \textsc{SeBa} galaxy model we checked whether the GC's right ascension and declination were within the 2$\sigma$ uncertainty range for those parameters for the DWD, and if so, considered it to be overlapping the GC. This is a loose criterion as it does not take into account the shape of the DWD's error ellipse (if it is very narrow, the GC could actually be far outside the 2$\sigma$ ellipse even if it is within 2$\sigma$ in latitude and longitude), but we chose to use this simple criterion for this filtering step for the sake of computational speed. The results on the number of overlapping sources for each GC are shown in Table \ref{overlap_table} and detailed in Sect. \ref{results_overlap_loc}.

Once we had obtained a set of overlapping DWDs for each GC, we defined a grid of points around the sky location of the GC. This grid contains 100 cells in a 10$\times$10 grid centered on the location of the GC, with a spacing of 1\degr between the cells. Using a grid with the same width (in terms of right ascension) for different GCs would result in the sky area of the grid being smaller for GCs with more polar sky locations; therefore, to keep the areas of the grids roughly consistent, we multiplied the width of each GC's grid by $\sec(\delta)$, where $\delta$ is the declination angle of the GC. In other words, for each GC, the grid cells are spaced by 1\degr in declination and $\sec(\delta)$\degr in right ascension. This $\sec(\delta)$ factor ranges from 1.0007 for M5 to 3.25 for 47 Tuc.

For each grid cell, we calculated the number of MW disc sources whose sky errors overlap the cell's location. We did this by, for each of the overlapping DWDs for each GC, computing a two-dimensional Gaussian distribution based on the sky location error ellipse from \textsc{GWToolbox}. We then calculated the probability density of this distribution at the four corners of each grid cell and averaged them to obtain a value that indicates the source's probability of being in that cell based on its sky location uncertainty. The sum of the probability values across all 100 cells in the 10\degr grid gives the probability that the source is anywhere within the grid, giving an indication of the degree to which that source can be said to ``overlap'' with the sky location of the GC. For example, a DWD located within the grid that has a very high S/N and thus a small error ellipse would give a value close to 1, whereas a DWD with an error ellipse many times larger than the grid would give a value close to 0.

The sum of the probability values for all the overlapping sources for a GC then gives an expected value of the number of MW disc DWDs ``close'' to the GC. This is shown in the rightmost column of Table \ref{overlap_table} and detailed in Sect. \ref{results_10deg}.

\subsection{Overlaps in distance error} \label{method_overlap_dist}

In the previous section, we evaluated the overlap between MW disc sources and the GCs based on the two-dimensional sky location. However, we also have information on a third dimension that could help differentiate MW disc sources and GC sources: their distance from Earth.

For each of the MW disc DWDs, we used \textsc{GWToolbox} to calculate their distance uncertainties, and used these values to compute a one-dimensional Gaussian distribution for each DWD. Then, for each GC, we summed the distance distributions of each MW DWD that overlaps with the sky location of the GC, weighted by multiplying the contribution from each DWD by its probability of being within the 10\degr grid around the GC as calculated in Sect. \ref{method_overlap_loc}. This produced an overall distribution of the distances of the overlapping MW DWDs, which we could then compare to the distance uncertainties of test sources placed in the GCs themselves.

\section{Results}

\subsection{Overlaps in sky location error} \label{results_overlap_loc}

\begin{table*}
    \centering
    \caption{Numbers of DWDs in the \textsc{SeBa} MW model whose sky errors overlap with the locations of the GCs.}
    \begin{tabular}{c c | c c c c c | c}
    \multirow{2}{*}{Index} & \multirow{2}{*}{Name} & \multicolumn{5}{|c|}{Overlapping within 2$\sigma$} & \multirow{2}{*}{Within 10\degr grid} \\
    & & All sizes & < 30\degr & < 10\degr & < 3\degr & < 1\degr & \\
    \hline
    (4) & 47 Tuc & 1 & 1 & 0 & 0 & 0 & 0.09 \\
    (11) & M5 & 1 & 1 & 0 & 0 & 0 & 0.18 \\
    (18) & NGC 362 & 3 & 2 & 0 & 0 & 0 & 0.28 \\
    (12) & M92 & 4 & 4 & 0 & 0 & 0 & 0.37 \\
    (8) & M13 & 5 & 5 & 0 & 0 & 0 & 0.68 \\
    \hline
    (19) & NGC 6752 & 17 & 10 & 1 & 0 & 0 & 2.12 \\
    (1) & $\omega$ Cen & 48 & 35 & 1 & 0 & 0 & 4.18 \\
    (7) & M14 & 434 & 433 & 9 & 0 & 0 & 27.4 \\
    (16) & NGC 6541 & 998 & 990 & 37 & 0 & 0 & 108 \\
    (20) & NGC 5927 & 610 & 589 & 102 & 0 & 0 & 177 \\
    \hline
    (14) & M9 & 2159 & 2157 & 275 & 3 & 1 & 266 \\
    (10) & M22 & 2465 & 2454 & 418 & 3 & 0 & 370 \\
    (5) & M19 & 2206 & 2203 & 345 & 7 & 0 & 374 \\
    (6) & M62 & 2476 & 2473 & 506 & 13 & 0 & 473 \\
    (15) & M28 & 3543 & 3533 & 926 & 16 & 0 & 773 \\
    (13) & NGC 6380 & 3661 & 3655 & 1157 & 35 & 1 & 1054 \\
    (9) & NGC 6440 & 4574 & 4572 & 1628 & 50 & 1 & 1246 \\
    (17) & NGC 6553 & 5258 & 5251 & 2137 & 107 & 7 & 1787 \\
    (3) & Liller 1 & 6149 & 6147 & 2902 & 188 & 6 & 2462 \\
    (2) & Terzan 5 & 6445 & 6442 & 3074 & 249 & 11 & 2587 \\

    \end{tabular}
    \tablefoot{The third column shows the number of LISA-detectable DWDs in the MW model that overlap with the sky location of each GC, as described in Sect. \ref{method_overlap_loc}. The fourth through seventh columns show the numbers if we impose an additional constraint that we only consider MW DWDs whose error ellipses are smaller than a size threshold. The final column shows the expected number of MW DWDs within the 10\degr grid around each GC, based on the sky location probability distribution of those DWDs.}
    \label{overlap_table}
\end{table*}

\begin{figure}
    \centering
    \includegraphics[width=\columnwidth]{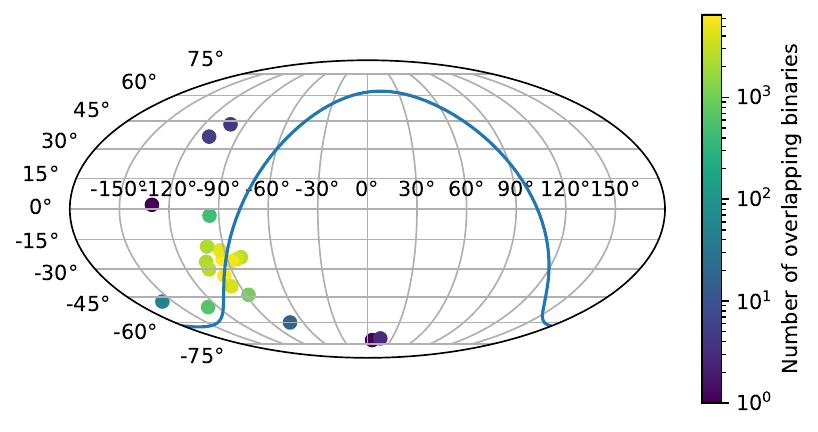}
    \caption{Sky locations of the 20 GCs in our sample. The GCs are coloured based on the number of overlapping DWDs from the MW disc. The blue line marks the Galactic plane.}
    \label{plot_gc_sky_loc}
\end{figure}

Table \ref{overlap_table} shows our results on the number of DWDs in the \textsc{SeBa} MW model whose sky errors overlap with the locations of the GCs. Specifically, the columns labelled ``overlapping within 2$\sigma$'' show the number of MW DWDs whose sky location uncertainties overlap with the location of the GC within 2$\sigma$ in both right ascension and declination, as described in Sect. \ref{method_overlap_loc}. The ``all sizes'' column is the total, and the other columns impose an additional restriction by only counting those DWDs whose error ellipses are smaller than a given threshold, e.g. the ``< 30\degr'' column only includes those sources whose uncertainty in both right ascension and declination is less than 30\degr.

It is apparent that there is a large variation in the number of overlapping DWDs for each GC, ranging from one for 47 Tuc and M5 to 6445 (roughly one-quarter of all resolvable DWDs in the MW model) for Terzan 5. The number of overlapping DWDs depends on the GC's sky location, as illustrated in Fig. \ref{plot_gc_sky_loc}: GCs far from the Galactic plane have only a few overlapping DWDs, while those close to the Galactic centre have thousands.

For most of the GCs, almost all of the DWDs that overlap with the GC's sky location have error ellipses smaller than 30\degr in both directions, but the majority are not below the 10\degr threshold. For GCs closer to the Galactic centre, DWDs with error ellipses smaller than 10\degr make up a larger proportion of the overlapping DWDs, which is to be expected: DWDs with small error ellipses will only overlap with the GCs if their sky locations are close together, which is more likely in regions of the sky with a higher density of DWDs.

\subsection{DWDs in the 10\degr grid} \label{results_10deg}

\begin{figure*}
    \centering
    \includegraphics[width=0.85\columnwidth]{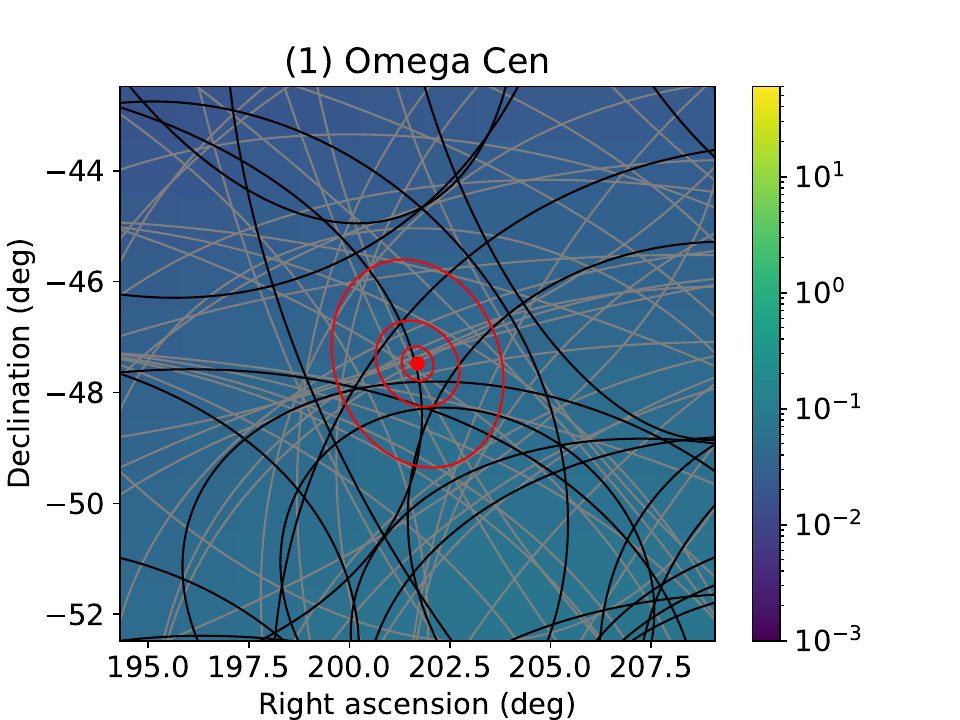}
    \includegraphics[width=0.85\columnwidth]{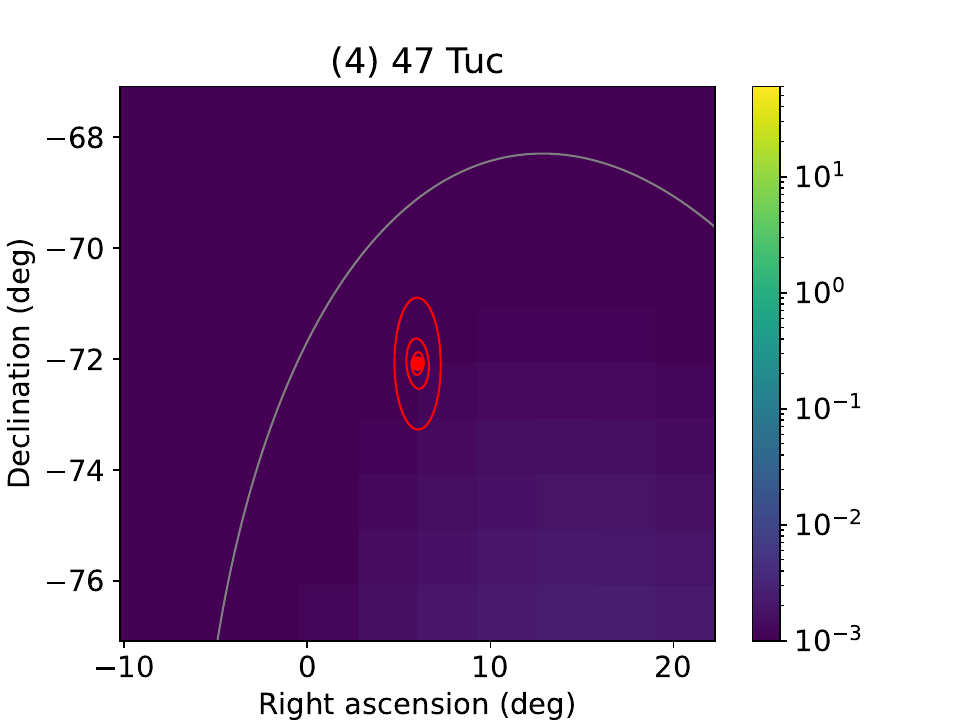}
    \includegraphics[width=0.85\columnwidth]{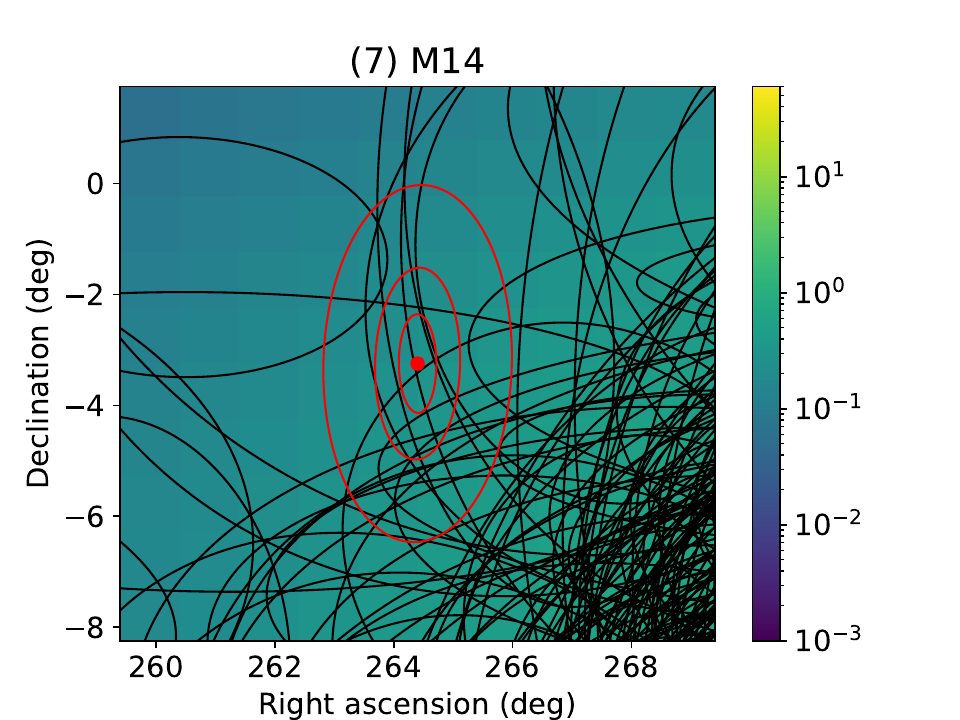}
    \includegraphics[width=0.85\columnwidth]{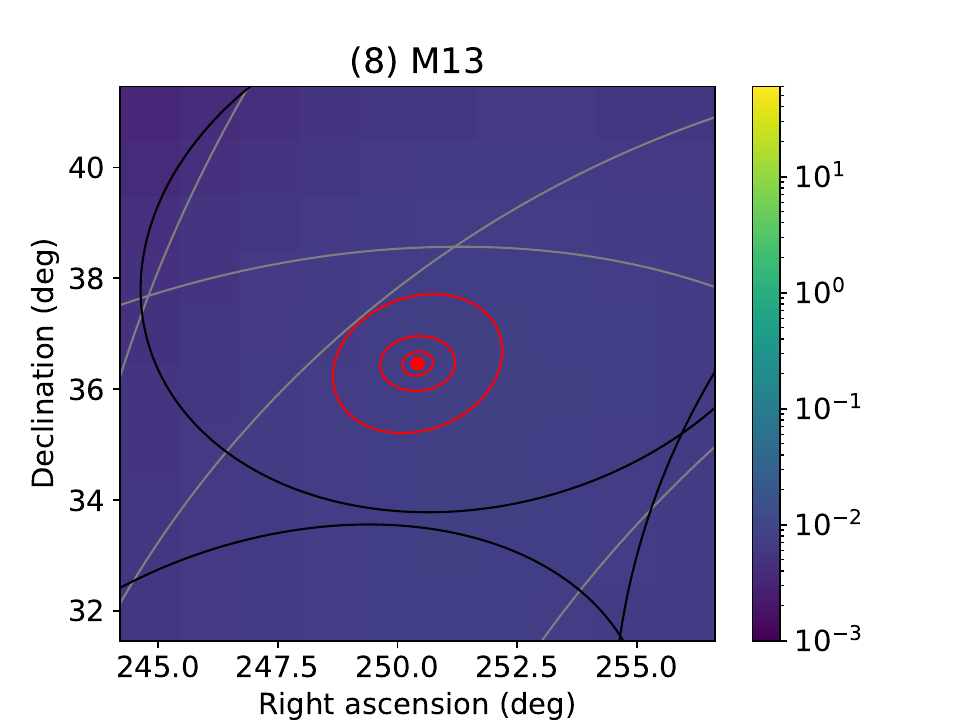}
    \includegraphics[width=0.85\columnwidth]{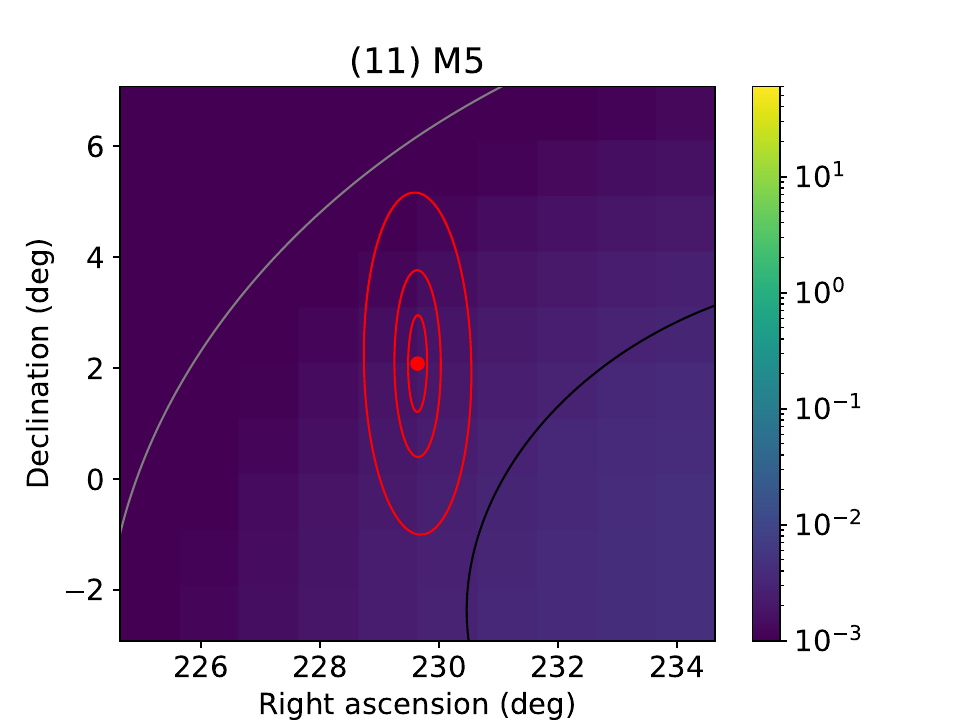}
    \includegraphics[width=0.85\columnwidth]{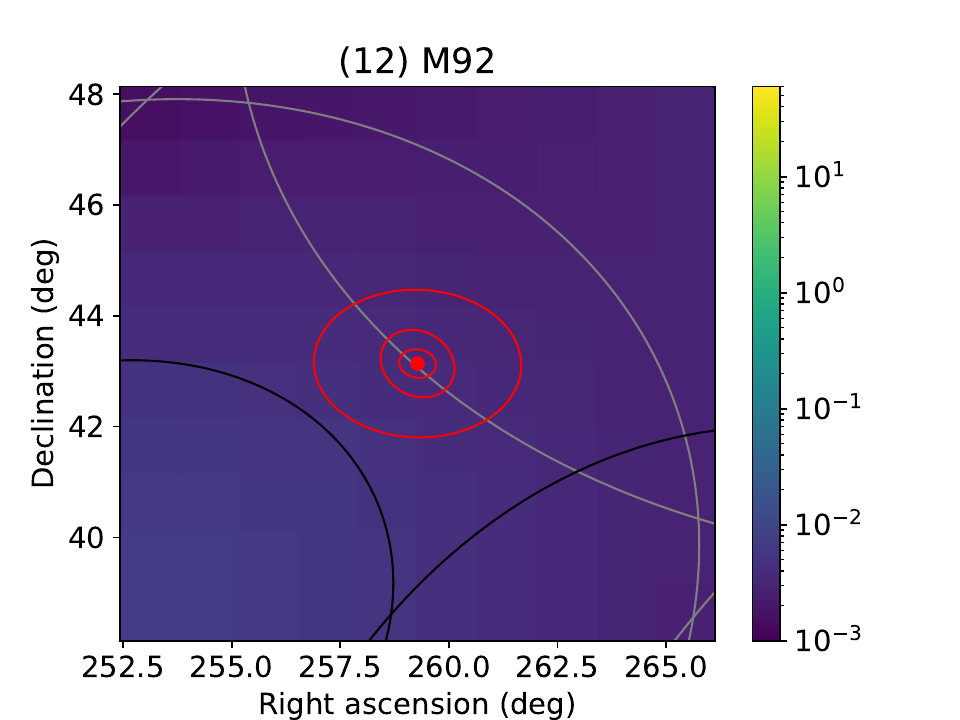}
    \includegraphics[width=0.85\columnwidth]{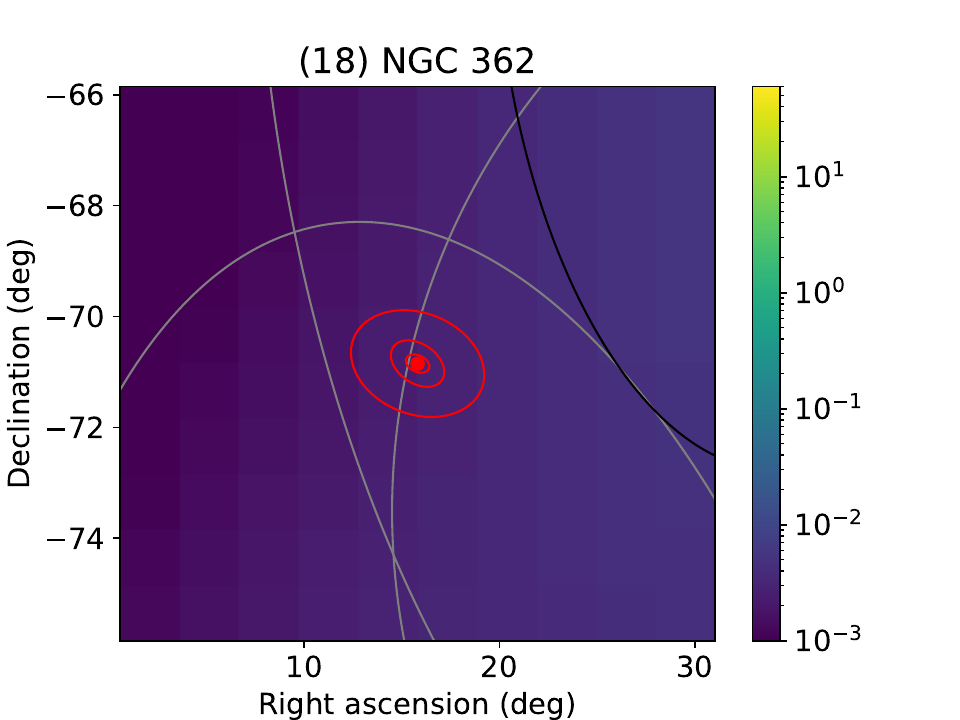}
    \includegraphics[width=0.85\columnwidth]{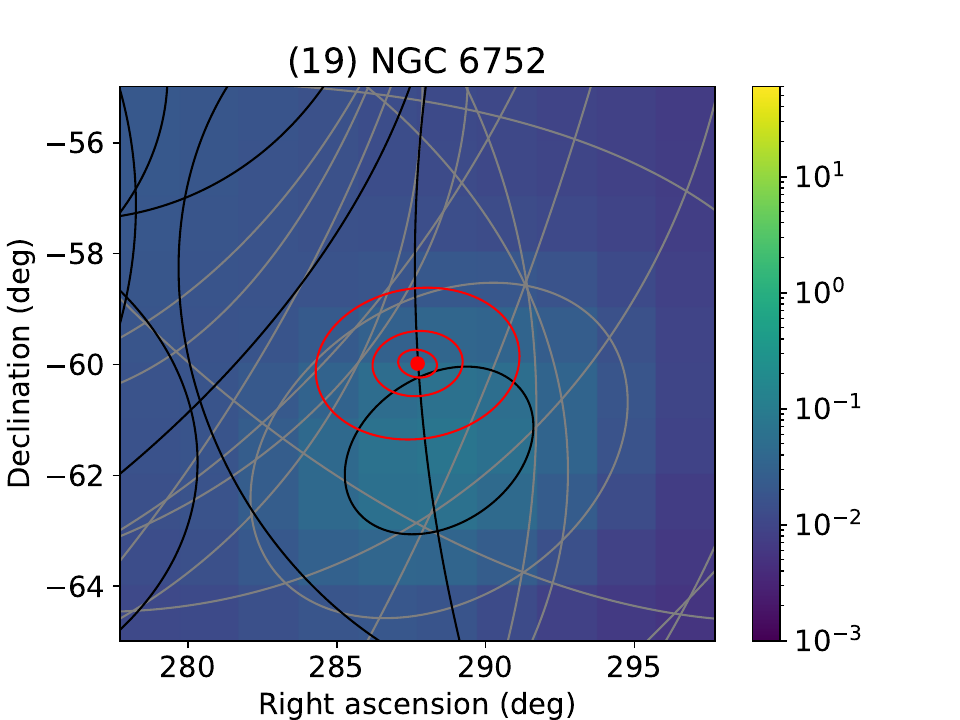}
    \caption{10\degr grid plots for GCs with <500 overlapping sources. The red dot marks the location of the GC, and the red ellipses are the 1$\sigma$ error ellipses for test binaries in the GC with S/N=10,20,40. The black ellipses are 1$\sigma$ for MW disc binaries, and the grey ellipses are 2$\sigma$. The 2$\sigma$ ellipses are omitted for M14 (panel 7) to reduce clutter. The background colours indicate the expected number of MW disc binaries in each cell based on their sky location uncertainty distributions.}
    \label{plot_10deg_low}
\end{figure*}

\begin{figure*}
    \centering
    \includegraphics[width=0.5\columnwidth]{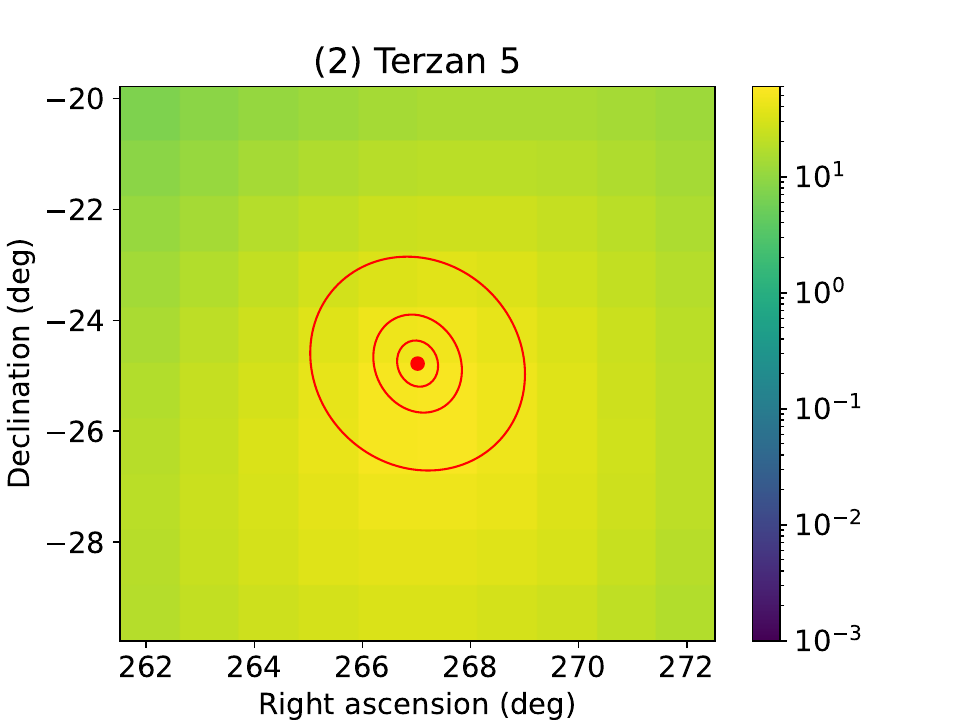}
    \includegraphics[width=0.5\columnwidth]{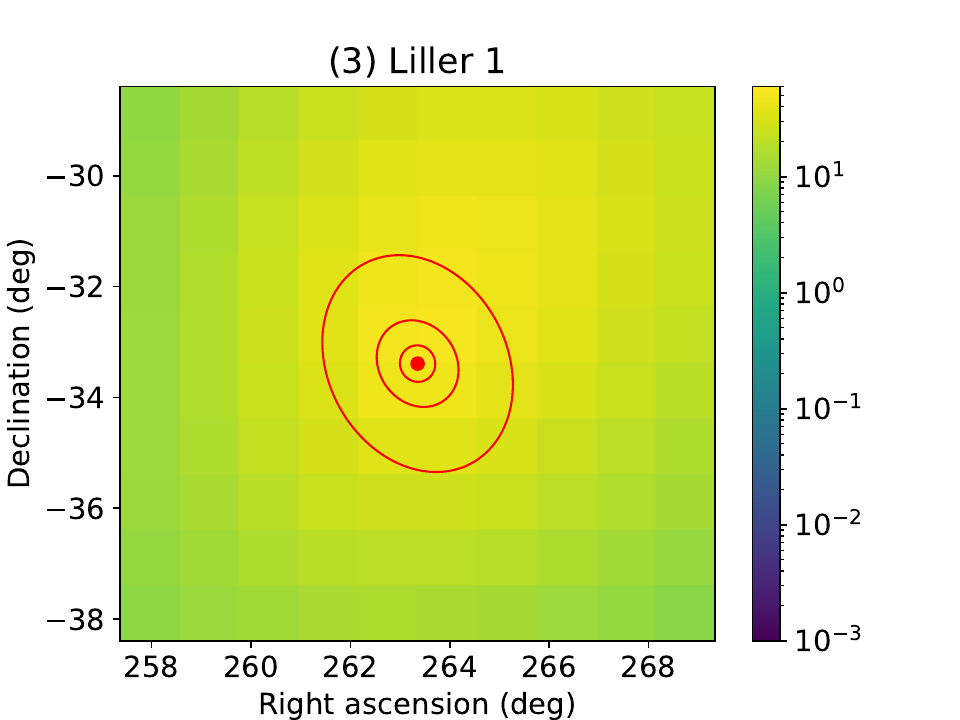}
    \includegraphics[width=0.5\columnwidth]{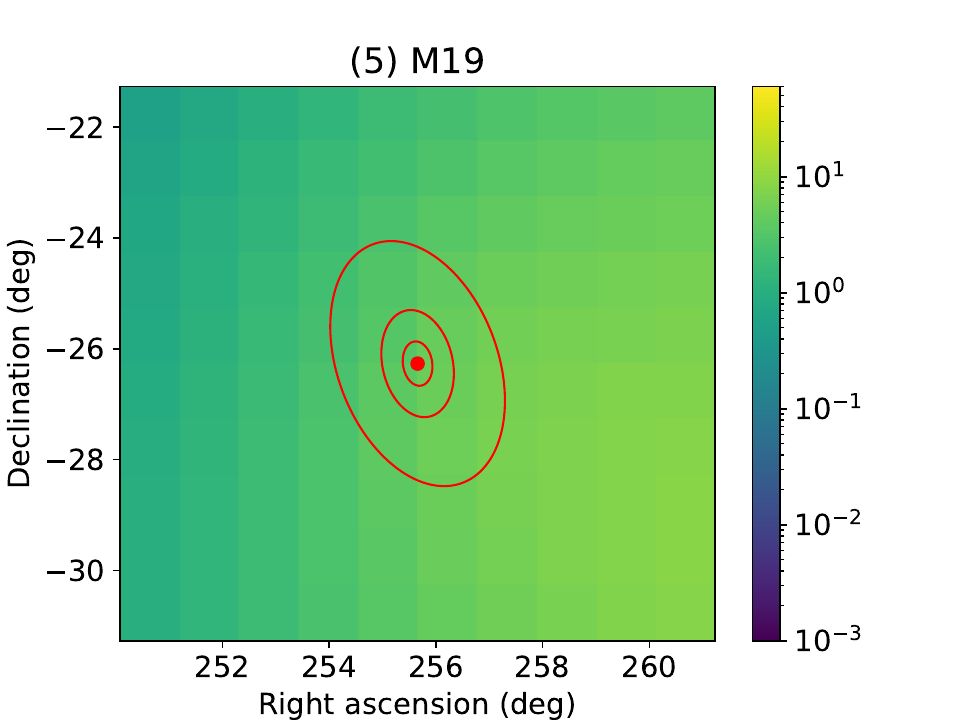}
    \includegraphics[width=0.5\columnwidth]{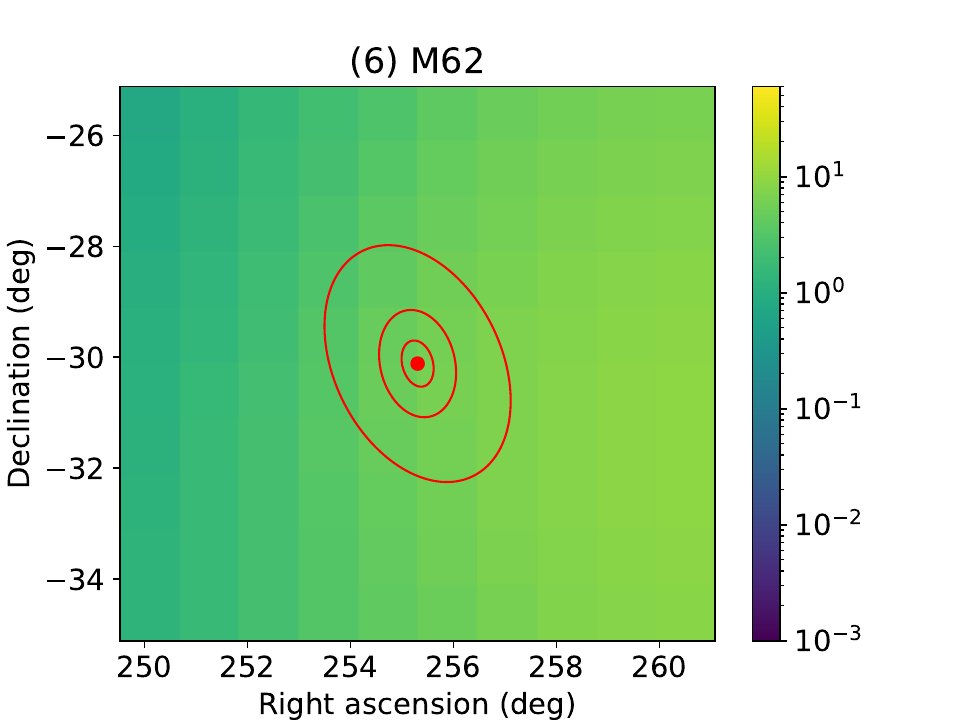}
    \includegraphics[width=0.5\columnwidth]{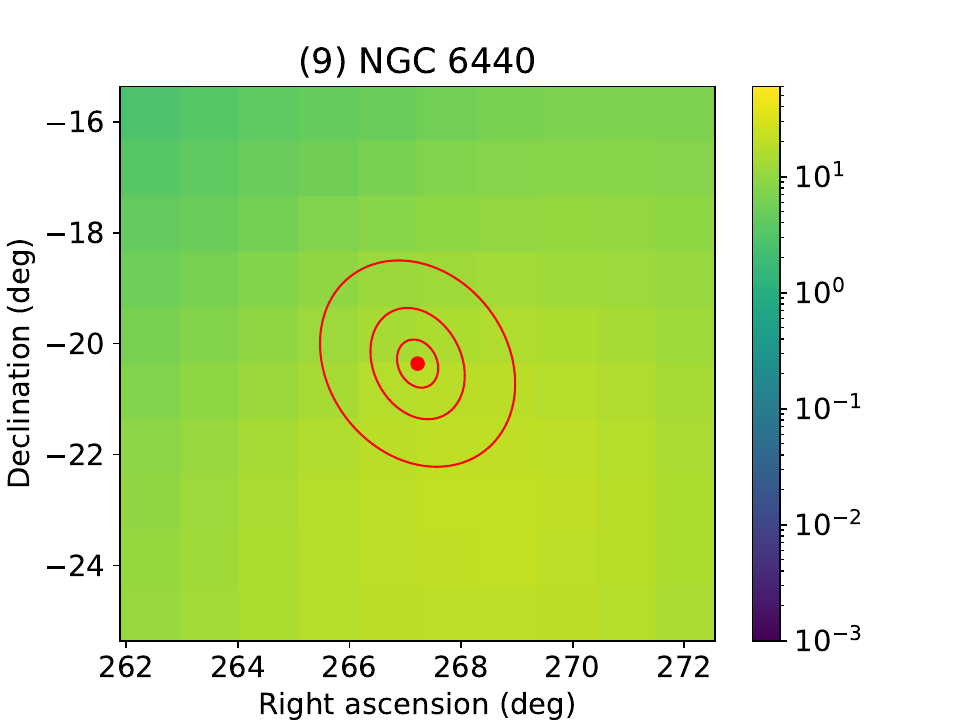}
    \includegraphics[width=0.5\columnwidth]{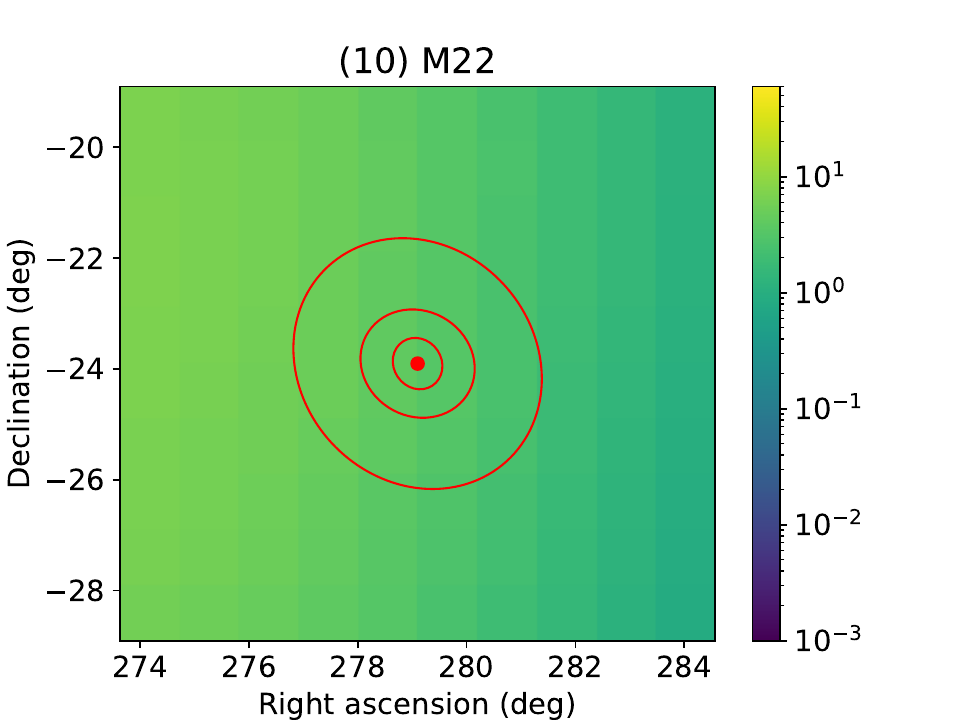}
    \includegraphics[width=0.5\columnwidth]{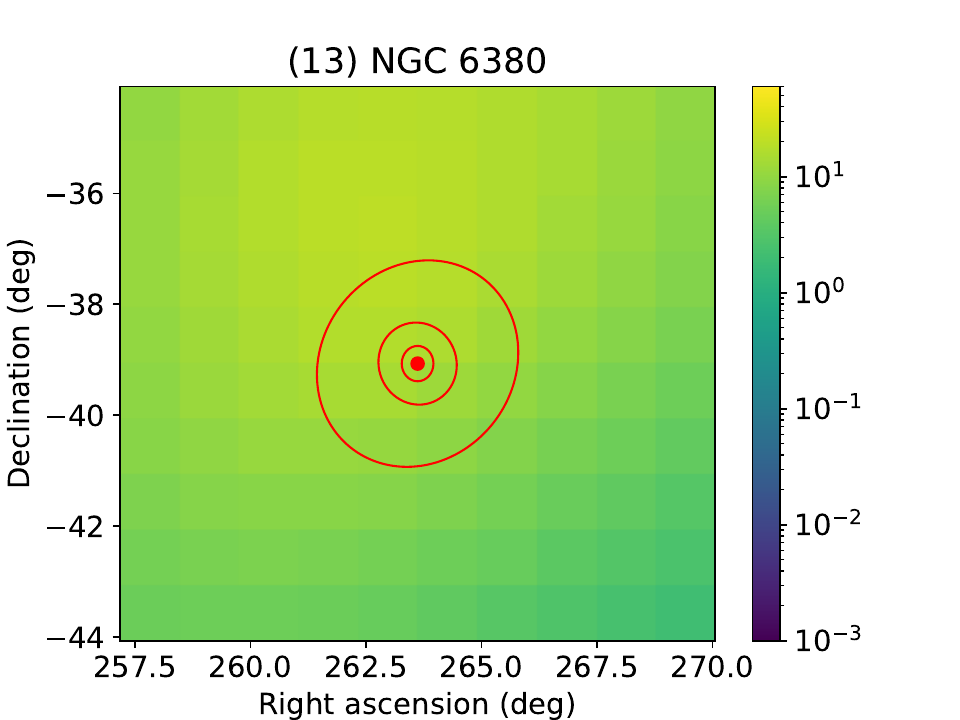}
    \includegraphics[width=0.5\columnwidth]{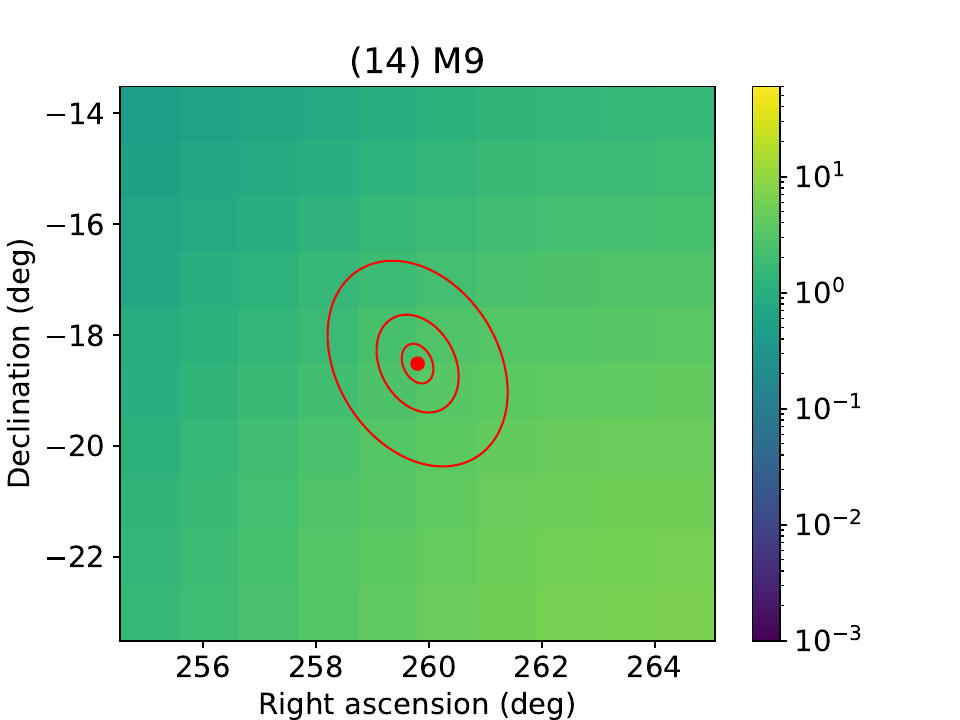}
    \includegraphics[width=0.5\columnwidth]{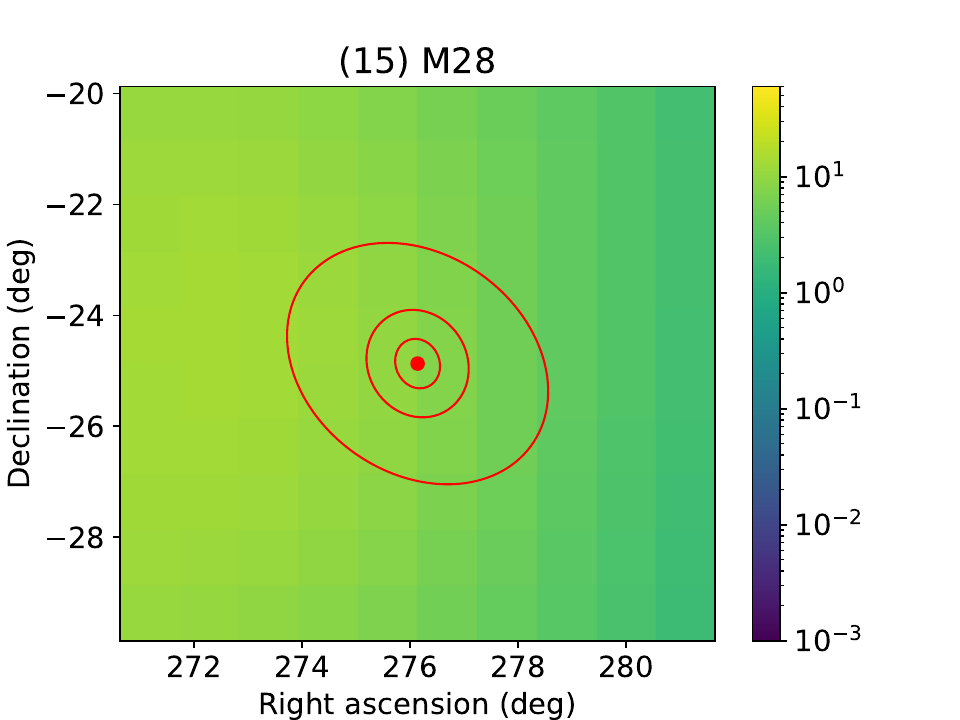}
    \includegraphics[width=0.5\columnwidth]{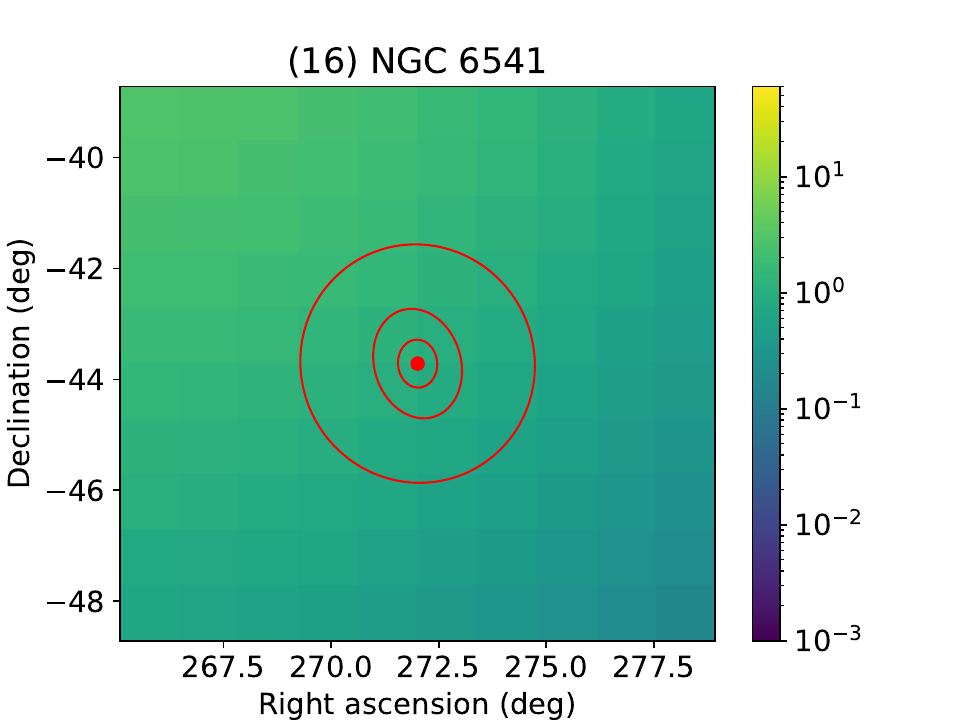}
    \includegraphics[width=0.5\columnwidth]{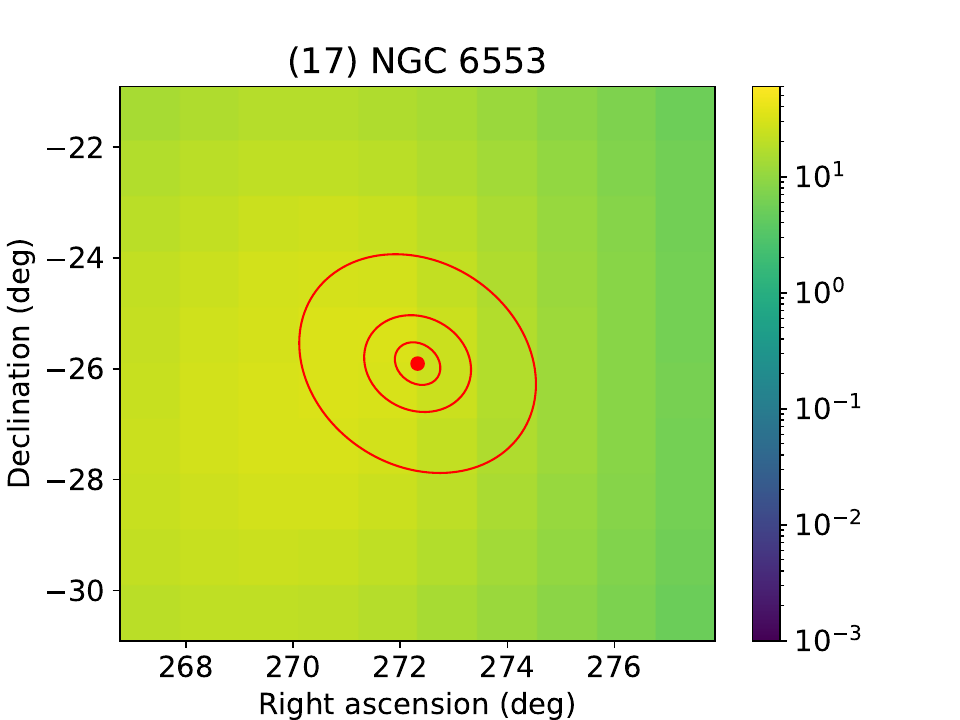}
    \includegraphics[width=0.5\columnwidth]{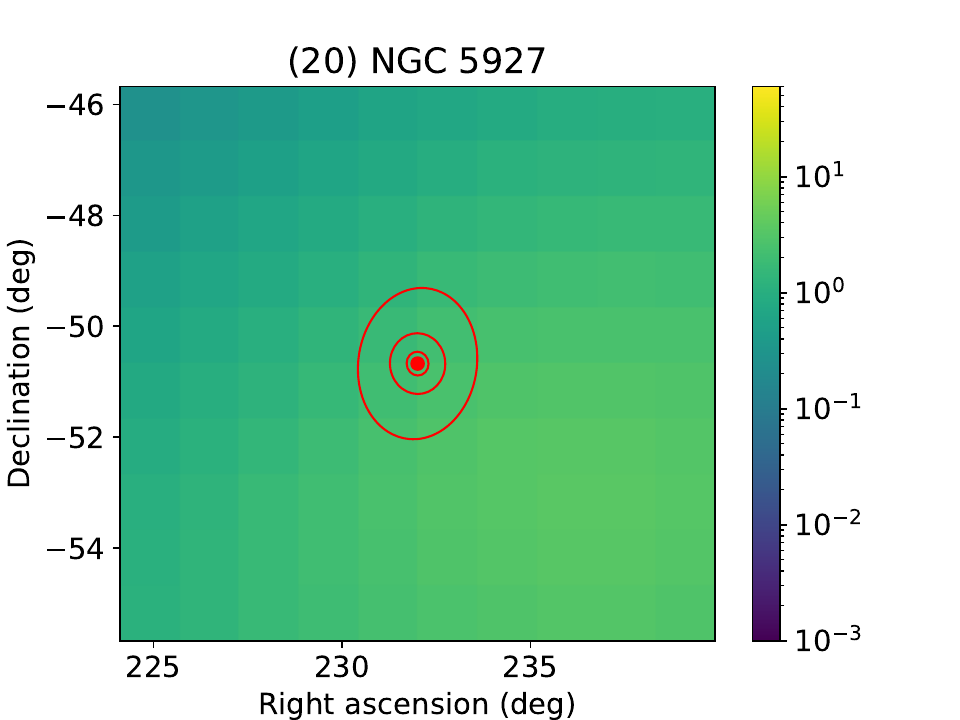}
    \caption{10\degr grid plots for GCs with >500 overlapping sources. Details are as for Fig. \ref{plot_10deg_low}, except the error ellipses of MW disc binaries are omitted to reduce clutter.}
    \label{plot_10deg_high}
\end{figure*}

In Figs. \ref{plot_10deg_low} and \ref{plot_10deg_high}, we show plots of the 10\degr grids around each of the GCs in our sample, as described in Sect. \ref{method_overlap_loc}. Fig. \ref{plot_10deg_low} shows the eight GC with less than 500 overlapping MW disc DWDs, and Fig. \ref{plot_10deg_high} shows the remaining 12 with more than 500.

The red dots mark the sky locations of each GC. The red ellipses are the 1$\sigma$ uncertainty regions of a test source placed at the location and distance of the GC. In these plots, we specifically use a NS–WD system with components of 1.35 and 0.6 M$_{\odot}$, respectively, and adjust its GW frequency between 1 and 5 mHz so that it matches a certain S/N threshold. From outward to inward, the three ellipses on each plot correspond to S/N values of 10, 20 and 40, respectively. As noted in Sect. \ref{method_test_sources}, for the same GC and S/N value, the different types of test binaries produced the same sky location uncertainties, so the specific masses we used here do not affect the result.

The black ellipses are the 1$\sigma$ uncertainty regions for the MW disc DWDs, and the grey ellipses are the corresponding 2$\sigma$ regions; for GCs with more than 100 overlapping MW DWDs we do not plot the 2$\sigma$ ellipses, and for those with more than 500 overlapping DWDs we do not plot any of the MW ellipses, in order to reduce clutter. The background colour of each 1\degr cell in the grid corresponds to the expected number of MW DWDs in that cell, obtained by summing the sky location uncertainty distributions of each DWD. As described in Sect. \ref{method_overlap_loc}, the rightmost column of Table \ref{overlap_table} shows the sum of these probabilities over the entire 10\degr grid, giving the expected number of MW disc DWDs in the 10\degr grid around each GC based on LISA's measurement uncertainties.

It can be seen that the expected number of MW DWDs in the 10\degr grid correlates with the total number of overlapping DWDs for each GC but is always smaller. This is expected, as the middle columns of Table \ref{overlap_table} show that many of the overlapping DWDs have error ellipses larger than the 10\degr grid, and so would contribute only a fractional to the sum within the grid. The values span more than four orders of magnitude: five of our GCs have less than one expected MW binary in the grid, and thus any source with significant S/N in the GC would be easily identified, while five other GCs have more than a thousand.

\subsection{Overlaps in distance error}

\begin{figure*}
    \centering
    \includegraphics[width=0.5\columnwidth]{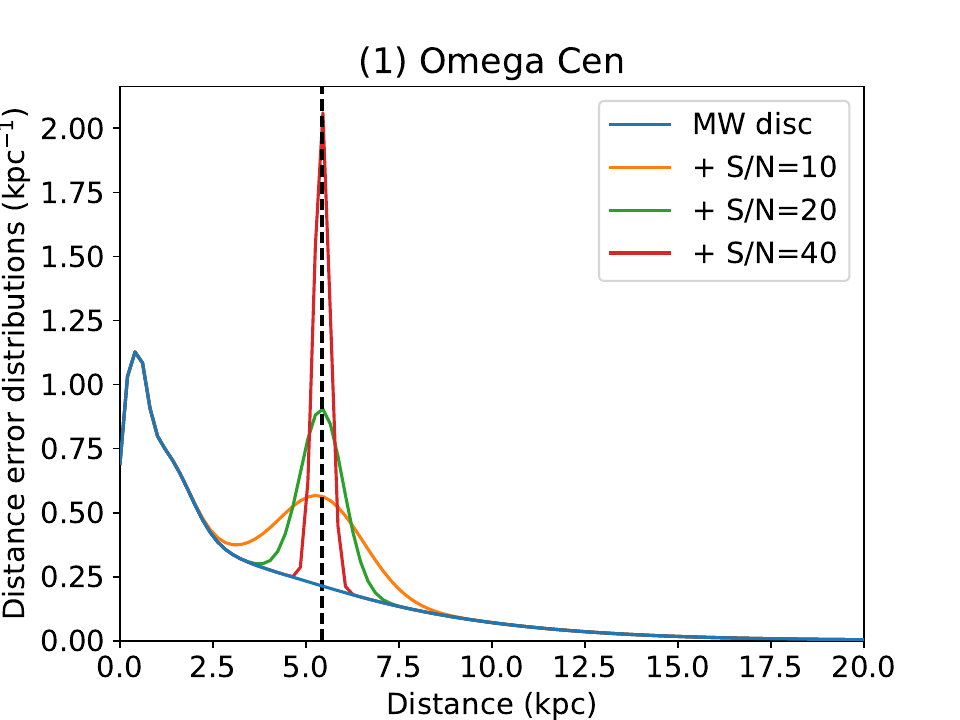}
    \includegraphics[width=0.5\columnwidth]{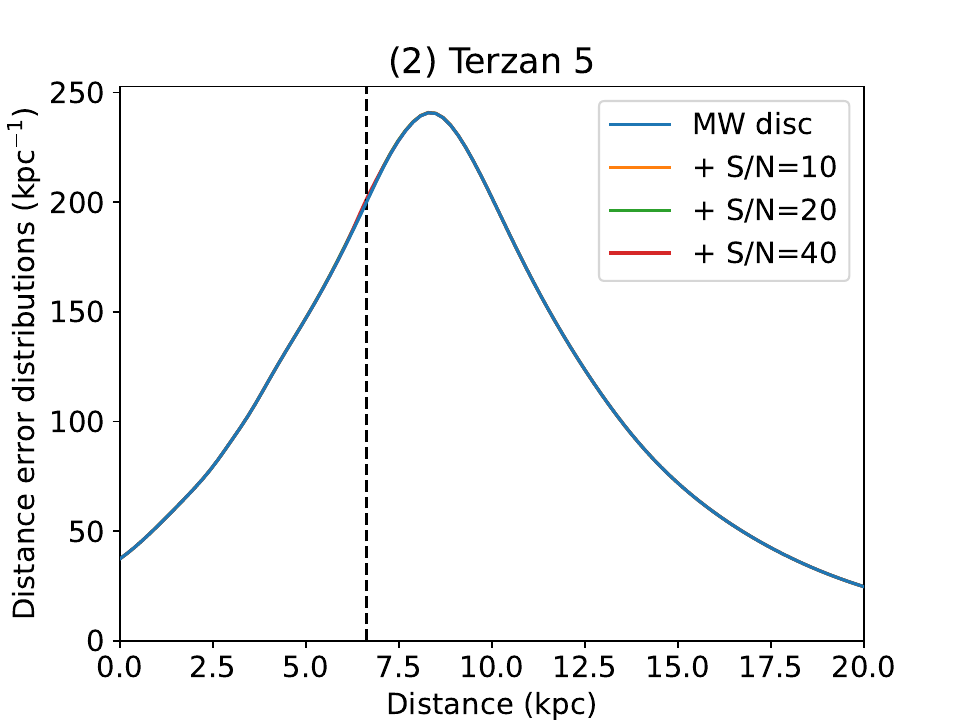}
    \includegraphics[width=0.5\columnwidth]{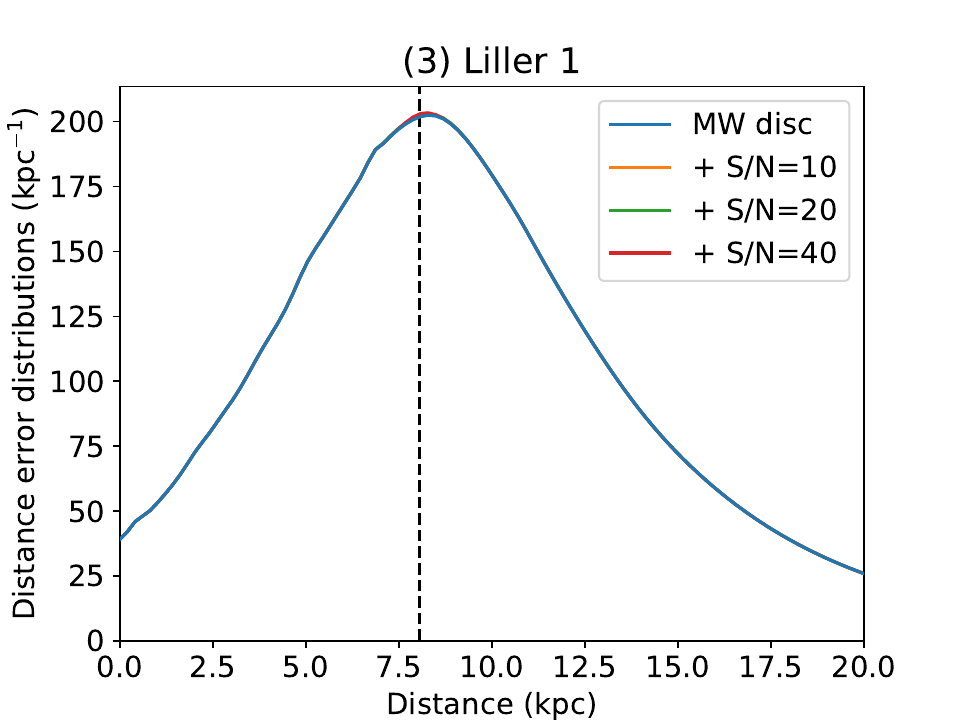}
    \includegraphics[width=0.5\columnwidth]{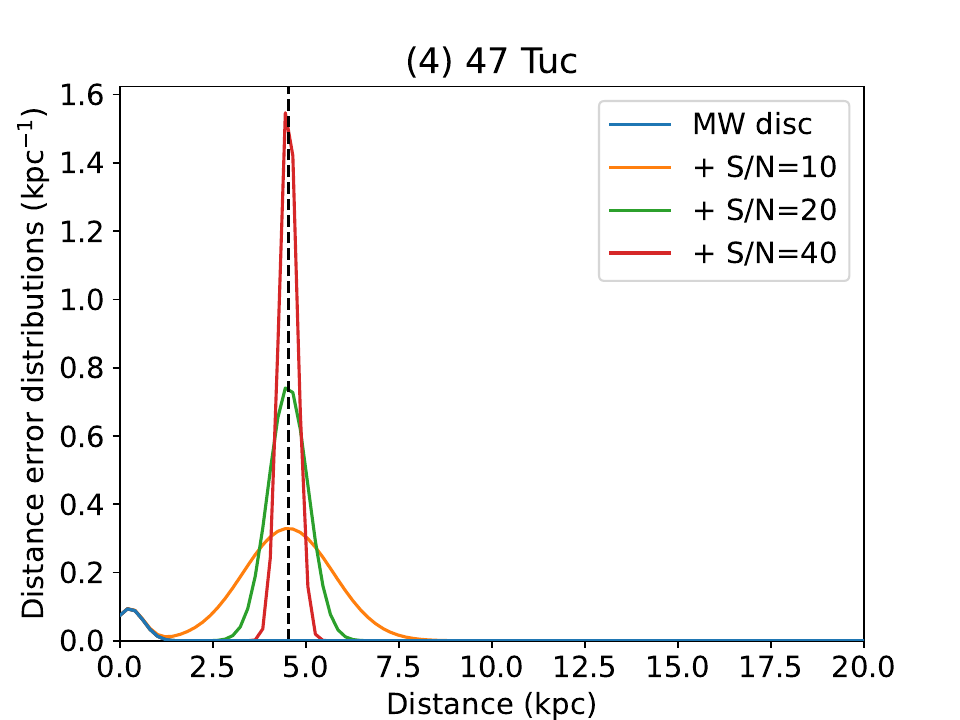}
    \includegraphics[width=0.5\columnwidth]{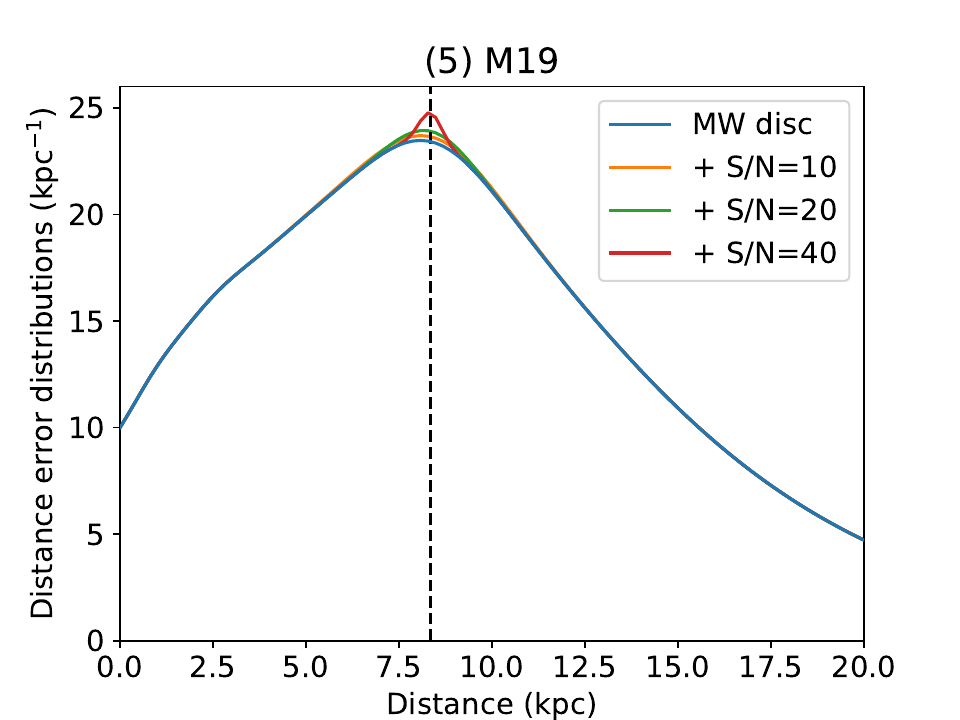}
    \includegraphics[width=0.5\columnwidth]{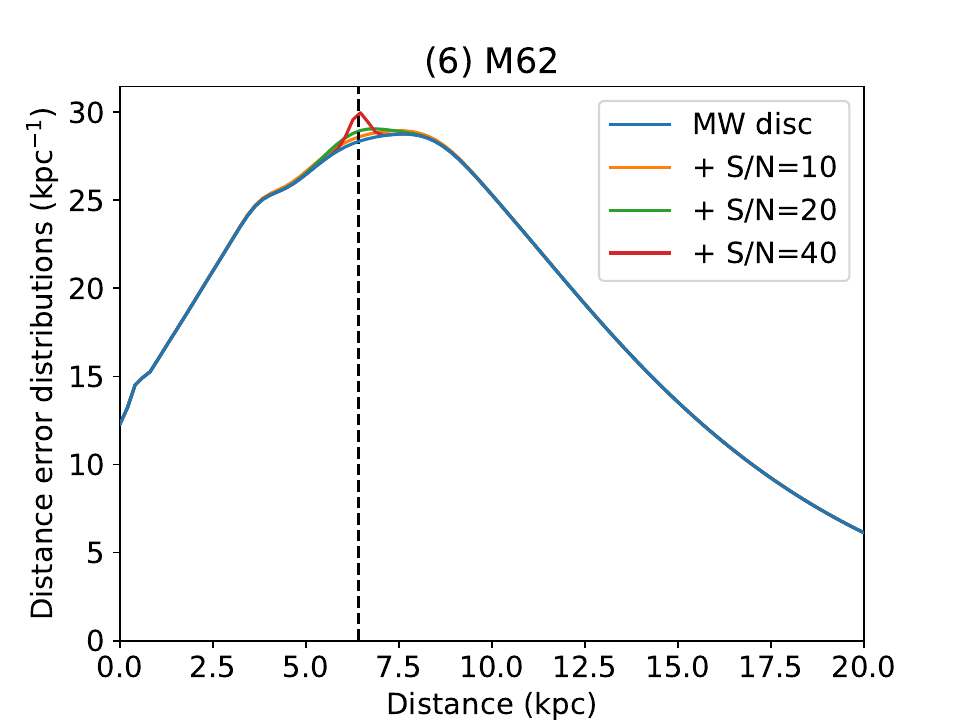}
    \includegraphics[width=0.5\columnwidth]{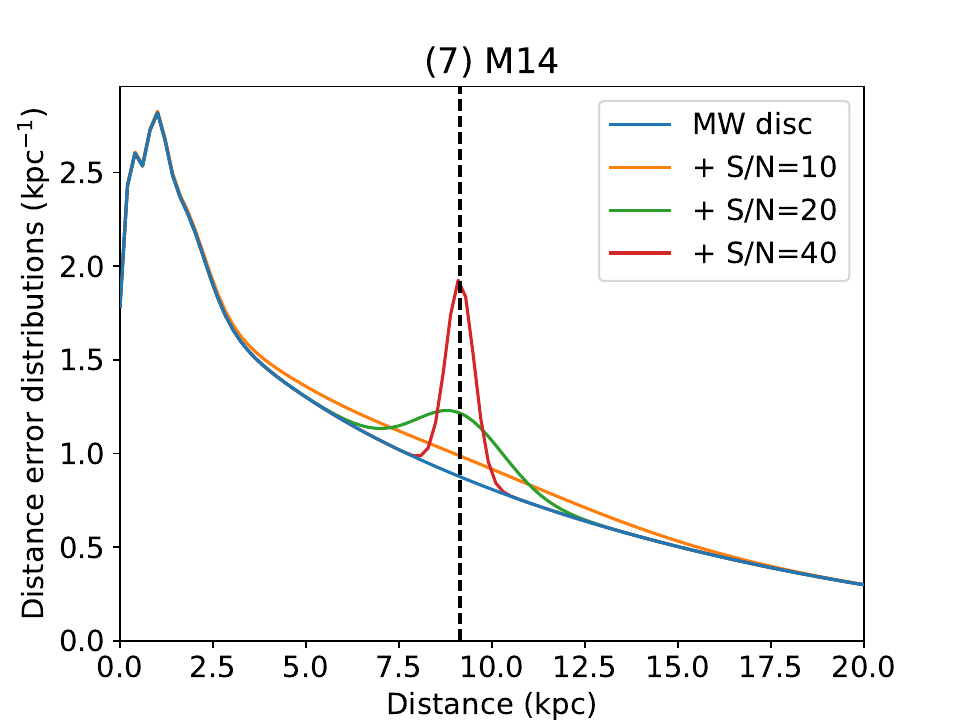}
    \includegraphics[width=0.5\columnwidth]{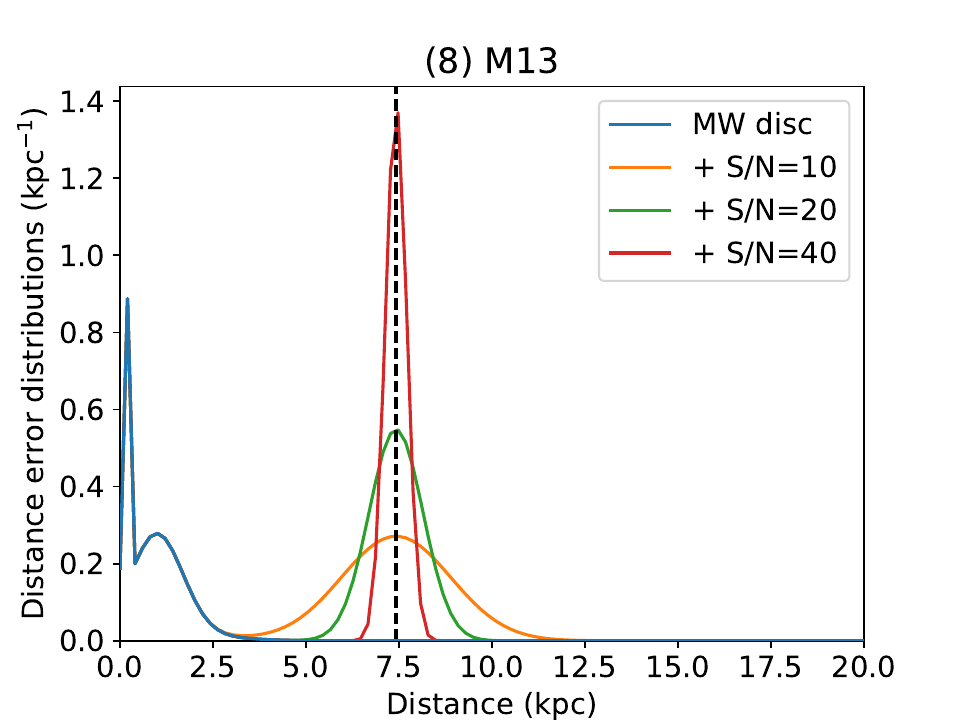}
    \includegraphics[width=0.5\columnwidth]{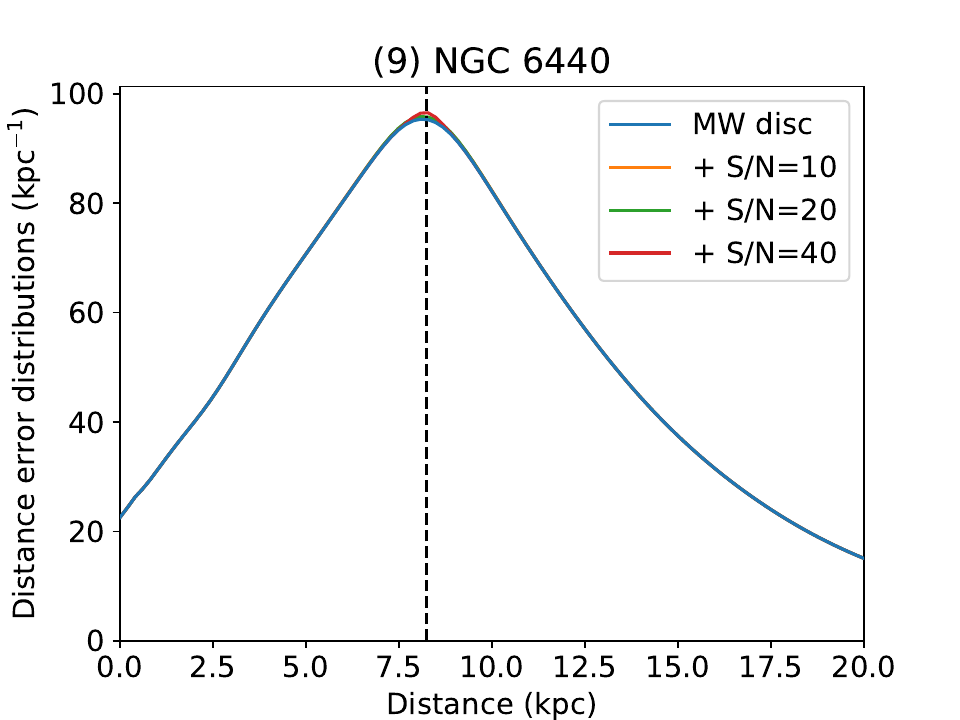}
    \includegraphics[width=0.5\columnwidth]{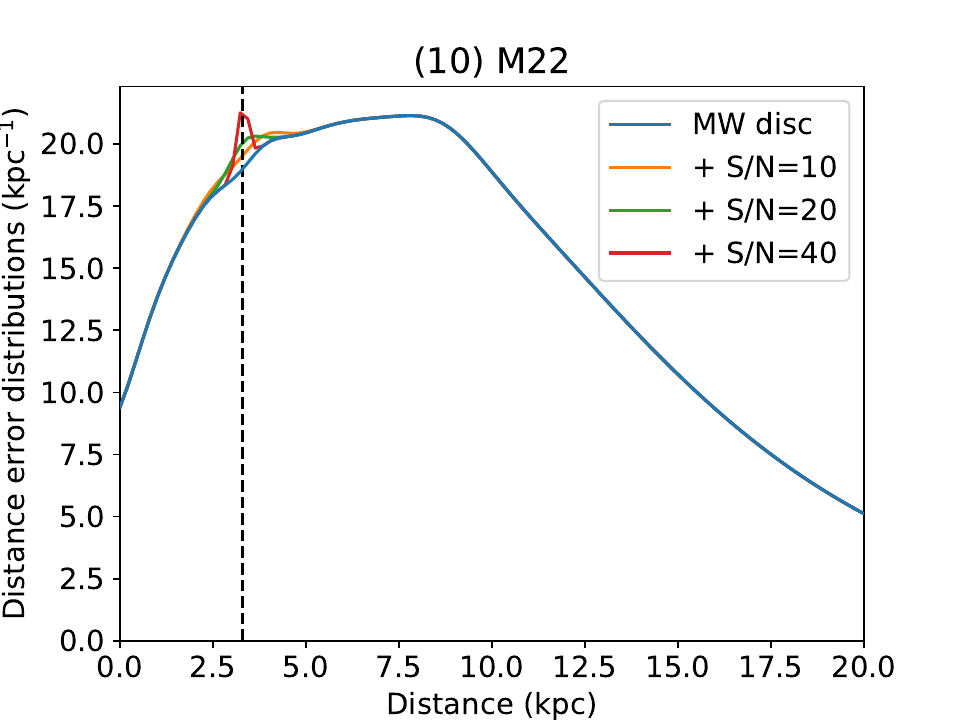}
    \includegraphics[width=0.5\columnwidth]{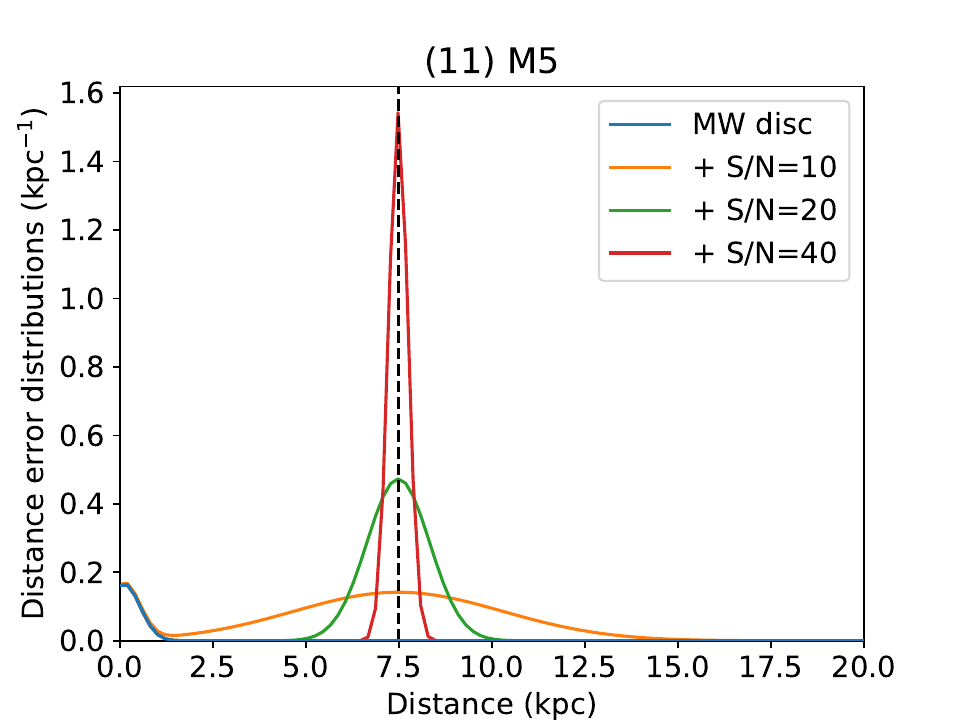}
    \includegraphics[width=0.5\columnwidth]{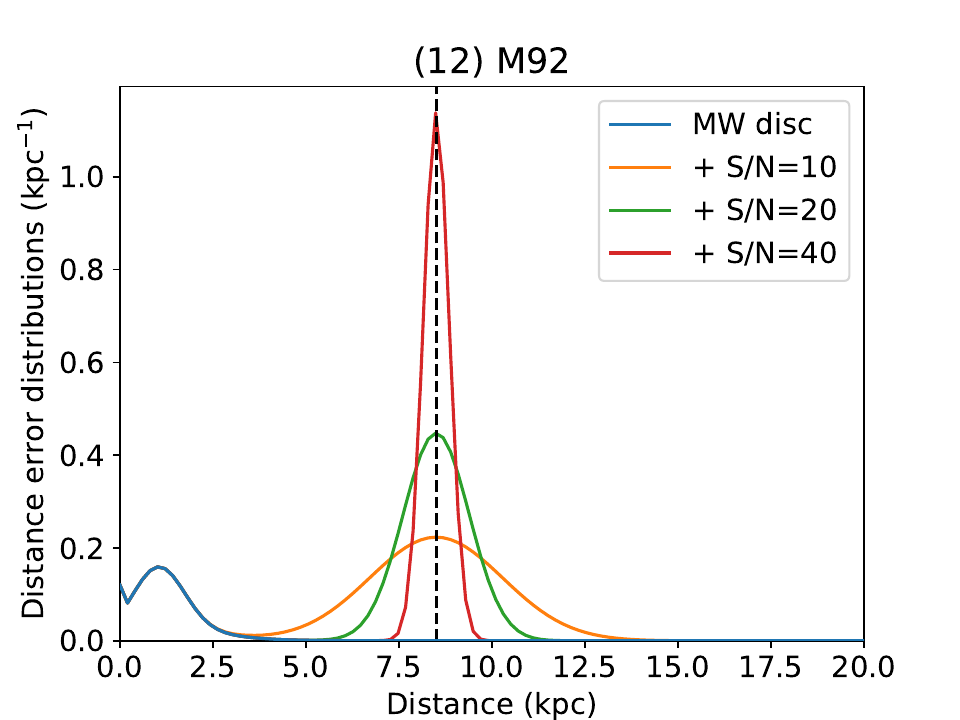}
    \includegraphics[width=0.5\columnwidth]{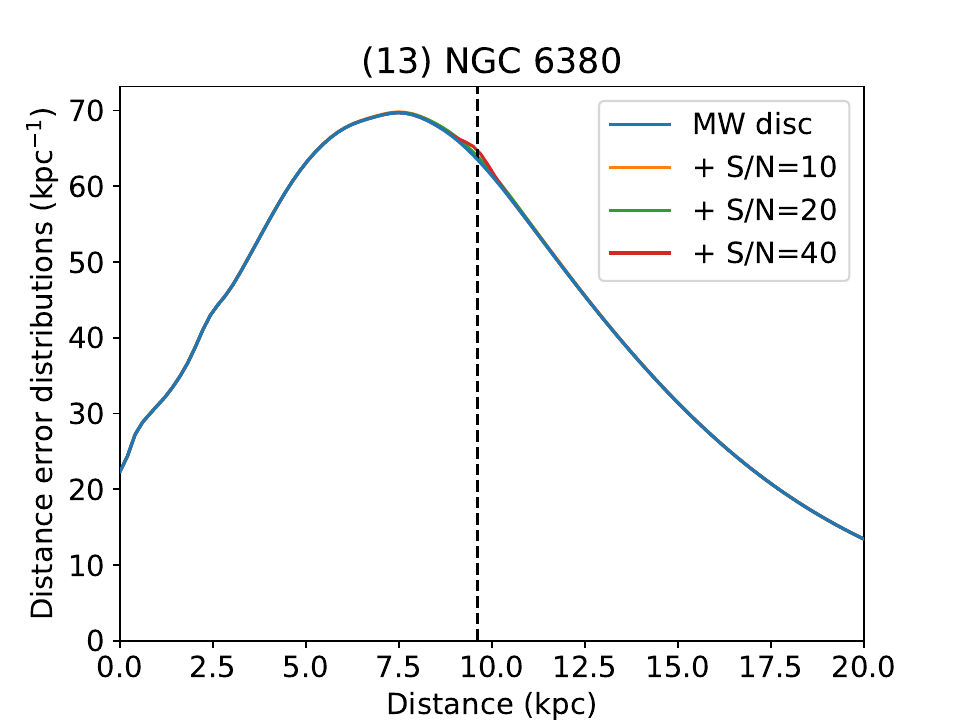}
    \includegraphics[width=0.5\columnwidth]{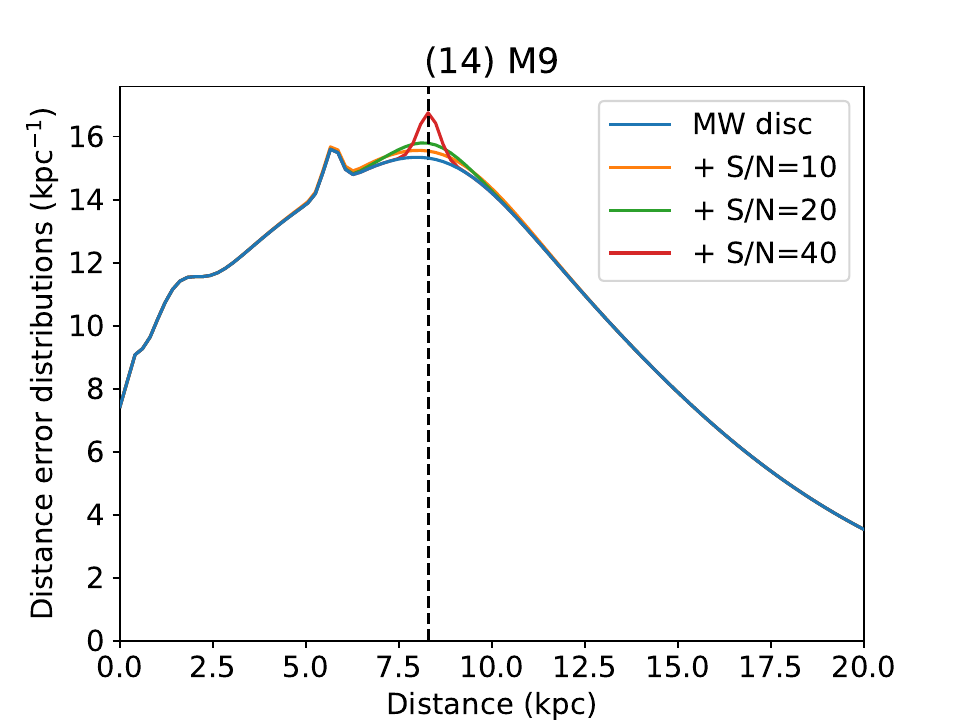}
    \includegraphics[width=0.5\columnwidth]{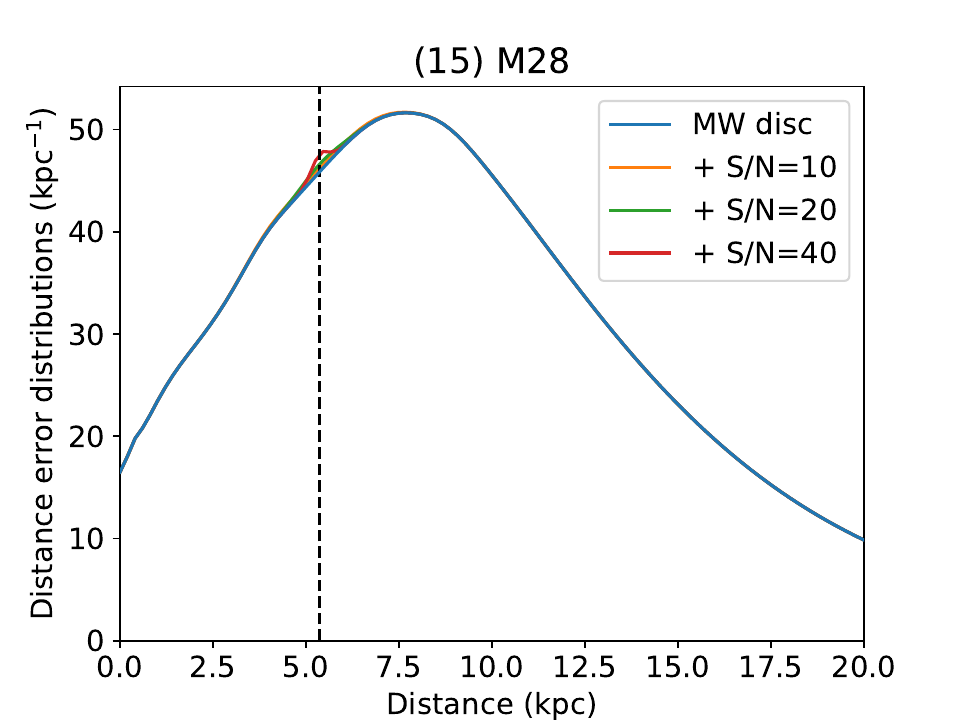}
    \includegraphics[width=0.5\columnwidth]{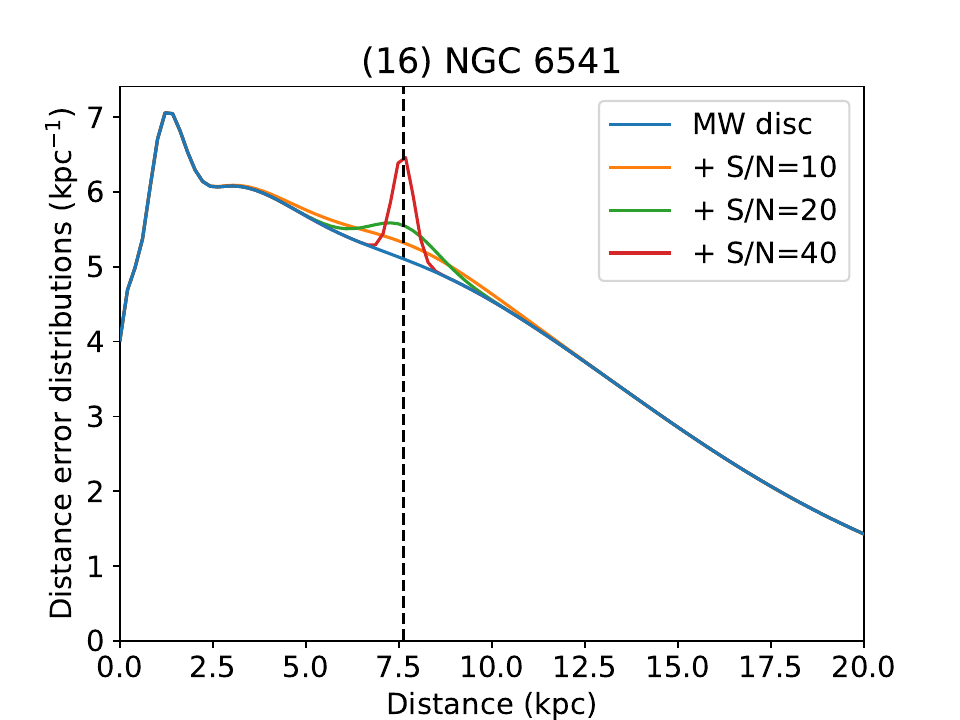}
    \includegraphics[width=0.5\columnwidth]{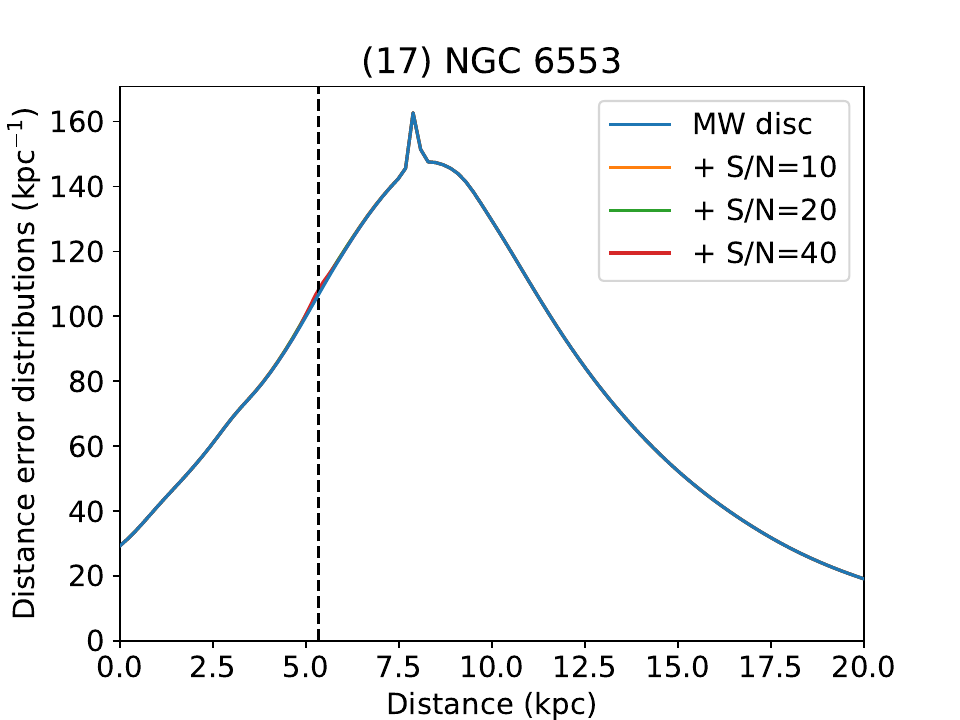}
    \includegraphics[width=0.5\columnwidth]{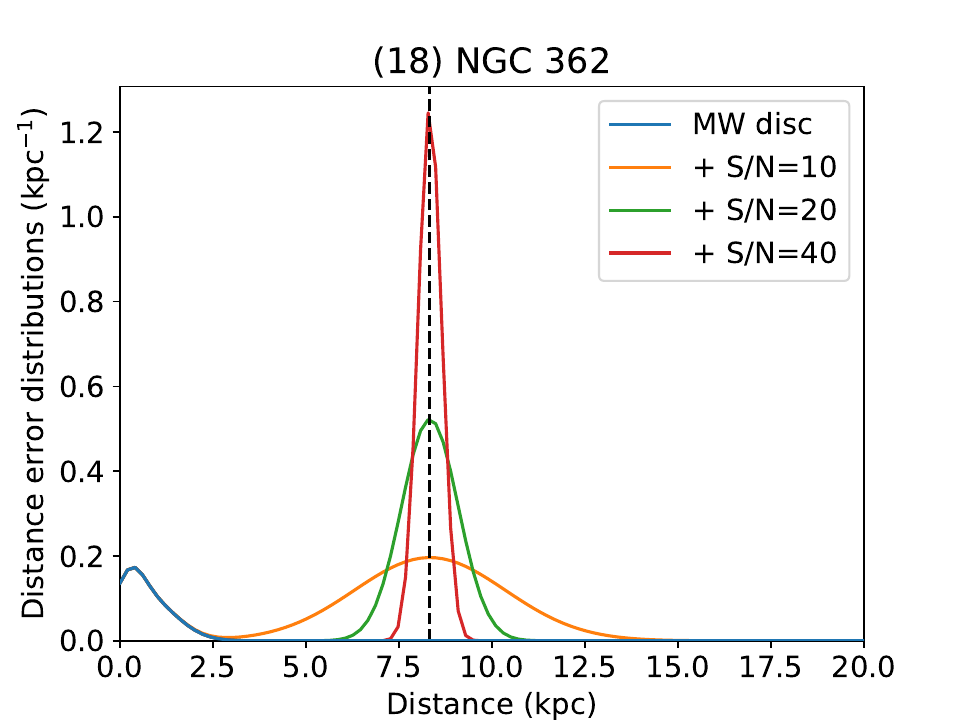}
    \includegraphics[width=0.5\columnwidth]{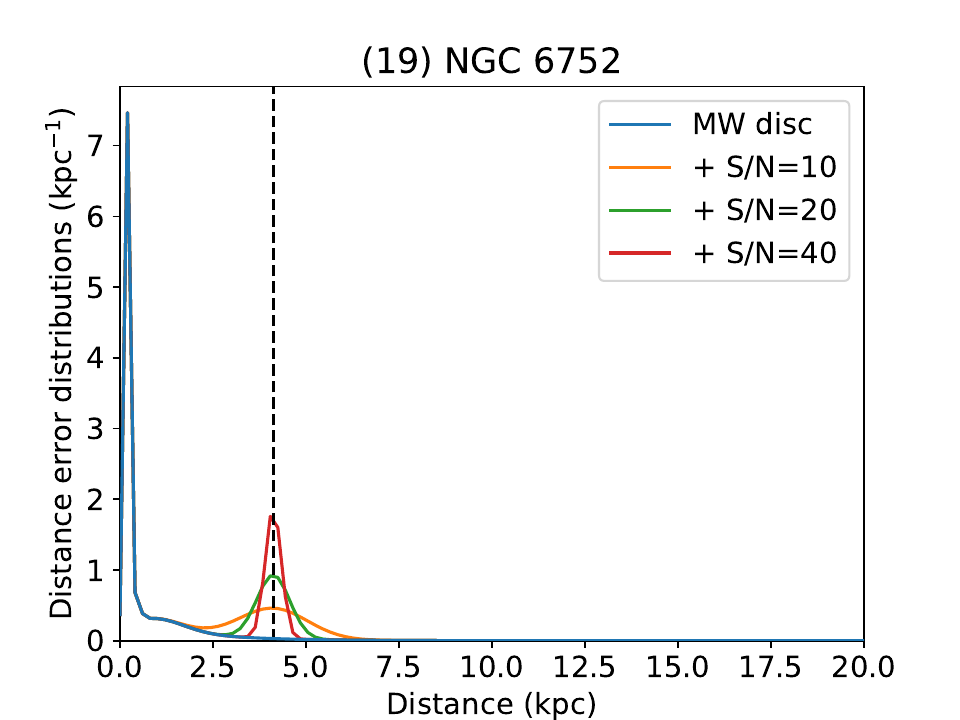}
    \includegraphics[width=0.5\columnwidth]{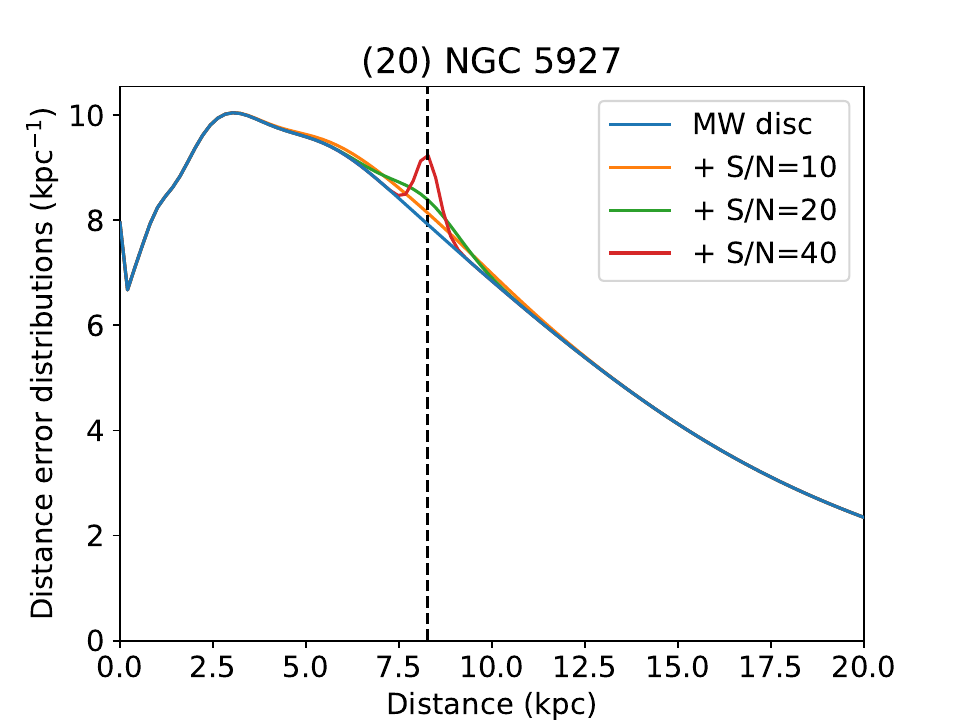}
    \caption{Comparison of distance error distributions between test binaries within a GC and the overlapping MW disc DWDs, for each of the 20 GCs in our sample. The blue lines are the weighted sums of the distance distributions of the overlapping MW DWDs for each GC. The other coloured lines add to this distribution a test binary at the distance of the GC with a certain S/N. The dashed black lines mark the distance of each GC.}
    \label{plot_dist_comp}
\end{figure*}

Finally, in Fig. \ref{plot_dist_comp} we compare the distance error distributions of test binaries within each GC to those of the overlapping DWDs from the MW disc. The blue lines show a weighted sum of the distance distributions of the overlapping MW DWDs, as described in Sect. \ref{method_overlap_dist}. For the other lines, we added on top of this ``background'' the distance error distribution for a single source in the GC. As in the previous section, we used test binaries with S/Ns of 10, 20 and 40.

We can see that, for those GCs with a very low number of overlapping binaries, such as 47 Tuc (panel 4) or NGC 362 (panel 18), the test binary in the GC can easily be distinguished from the MW DWDs even if the S/N is just 10, as one would expect. Not only are there few overlapping binaries, but the GC is also distinctly farther away from Earth than those MW binaries. The latter is caused by the fact that, when we ``look'' at a GC outside of the Galactic plane, there is only a short cross-section of the Galactic plane along our line of sight.

Conversely, for those GCs located close to the Galactic centre, such as Liller 1 (panel 3) or NGC 6440 (panel 9), the test binary in the GC is indistinguishable among the thousands of overlapping MW DWDs. Furthermore, since these GCs are close to the Galactic centre not only in sky location but also in distance, LISA's measurement of the distance to the GC binary will not help in distinguishing it from the MW disc binaries, since the distance to the GC coincides with the peak of the distance distribution of the MW disc DWDs.

There are also intermediate GCs that are relatively near the Galactic plane but not immediately adjacent to the Galactic centre, such as $\omega$ Cen (panel 1), M14 (panel 7) and NGC 6541 (panel 16). For these GCs, overlapping MW DWDs are present, but the GC is located at a larger distance than the peak of the distance distribution of the MW DWDs. Consequently, the test binary in the GC does stand out in the distance distribution, although only if the S/N is relatively high (which is easier to achieve with a NS-WD or heavy DWD than a light DWD).

Combining our insights from the sky location and distance uncertainty comparison, we can divide the 20 GCs in our sample into three classes, based on how well LISA could identify binaries within those GCs as belonging to those GCs. These classes are shown by the three subdivisions in Table \ref{overlap_table}. The first grouping contains five GCs (47 Tuc, M5, NGC 362, M92 and M13) which have fewer than one expected MW disc binary within a 10\degr grid around their sky location, and so any LISA-detectable binaries within these GCs could easily be identified as belonging to the GCs based on LISA's measurements of their sky locations. The second grouping contains five GCs (NGC 6752, $\omega$ Cen, M14, NGC 6541 and NGC 5927) which have more overlapping MW disc DWDs, and so resolving binaries to these GCs may be difficult based on the sky location information alone, but the additional information from LISA's distance measurements should make it possible to identify the binaries in the GCs. Finally, the third grouping contains ten GCs (M9, M22, M19, M62, M28, NGC 6380, NGC 6440, NGC 6553, Liller 1 and Terzan 5) that are located close to the Galactic centre and therefore have so many overlapping MW disc DWDs that binaries within the GCs could not be distinguished from them, even when considering LISA's measurements of both their sky location and distance.

Considering the number of overlapping MW disc binaries for each GC as well as the GCs' masses, the most promising GC for finding a WD binary that LISA could resolve to that GC is 47 Tucanae (panel 4 in Figs. \ref{plot_10deg_low} and \ref{plot_dist_comp}), which is located far away from the Galactic plane, having the lowest number of overlapping binaries in our MW model, and is one of the most massive GCs. Omega Centauri (panel 1 in Figs. \ref{plot_10deg_low} and \ref{plot_dist_comp}), the most massive GC orbiting the MW, does have some overlapping MW disc binaries, but using LISA's sky location and distance measurements together, it should be possible to resolve any LISA-detectable binaries in Omega Centauri to that GC, particularly if they have a S/N of 20 or above.

\section{Discussion}

When comparing the binaries in GCs with those in the MW disc, we placed a single binary in each GC. As mentioned previously, predictions of the number of LISA-detectable binaries in GCs vary significantly, but for those predictions on the higher end \citep[e.g.][]{willems2007,hellstrom2025} it is possible that there may be multiple LISA-detectable DWDs in a GC, particularly for more massive GCs. If multiple binaries were to be present in a GC, it would be easier to detect that GC as an overdensity compared to the MW disc in terms of sky location and distance. This is particularly relevant for the second class of binaries we identified, as the first class are already identifiable with a single binary, and the third class would require tens or hundreds of binaries in a GC to be distinguishable from the MW disc. However, identifying any individual binary as being part of the GC or not would not necessarily become easier.

For our MW model, we used a single realisation of a simulation of the MW. The locations of binaries in the model are semi-random and so do not correspond directly to binaries in the real MW. For a more accurate comparison, one could use multiple realisations of a MW model and then average the results for each of these realisations. However, this would require a large amount of computational time and, given the many orders of magnitude difference in the number of overlapping DWDs between different GCs, we do not think this would significantly affect our conclusions.

As mentioned previously, the sky location and distance errors calculated by \textsc{GWToolbox} are not identical to those calculated by the LISA Figures of Merit effort in \citet{lisa_redbook}. The GW S/N values from \textsc{GWToolbox} are also not identical to those calculated by the alternative software package \textsc{legwork} \citep{legworkjoss,legworksci}, although \textsc{legwork} cannot perform error estimations. We also note that the Fisher matrix analysis method, used by both \textsc{GWToolbox} and \citet{lisa_redbook}, has some limitations when applied to GW parameter estimation, compared to more computationally intensive Monte Carlo simulations \citep{vallisneri2008}. However, as with the previous point, we do not think that the differences between software packages would affect our conclusions compared to the large differences between GCs.

\subsection{Core-collapsed GCs}

We selected the GCs in our sample based on their mass and proximity to Earth, since these factors make it more likely that a LISA-detectable binary would be found in these clusters as opposed to other GCs. However, a distinction can also be made between GCs by whether or not they have undergone core collapse, a phenomenon in which an instability in the gravitational equilibrium of a GC causes many stars to sink to its centre, increasing the central density of the GC \citep[see e.g.][]{henon1961,lynden1968,cohn1980}. About one-fifth of the roughly 150 GCs around the MW exhibit a density profile consistent with core collapse \citep{djorg1986,harris2010}.

There have been studies based both on observations \citep[e.g.][]{vitral2022} and on GC dynamics simulations \citep[e.g.][]{kremer2021} that have suggested that core-collapsed GCs have more numerous central populations of WDs than non-core-collapsed ones, and core-collapsed GCs are therefore more likely to host compact WD binaries. Furthermore, studies such as \citep{pooley2003} have predicted that the number of compact binaries in a GC scales with its central density.

Consequently, nearby core-collapsed GCs such as NGC 6397 \citep{kremer2021,vitral2022} could also be potential hosts of LISA-detectable WD binaries, despite the fact that they are less massive than the GCs in our sample. However, we do not think that adding such GCs to our sample would affect our results on the relation between a GC's sky location and the probability that LISA would be able to resolve a DWD in that GC to that GC.

\subsection{Eccentricity in GC DWDs} \label{discussion_eccentricity}

The \textsc{SeBa} MW model we used contains only DWDs with circular orbits (i.e. zero eccentricity), and the GC test binaries we used are also all circular. This is generally a reasonable assumption for DWDs in the MW disc, as DWDs with orbits tight enough to be LISA-detectable would have gone through at least one common-envelope event, which leads to circularisation of the binary.

However, if a DWD were to have an outer tertiary companion, this could increase the eccentricity of the inner binary \citep{naoz2013,antognini2014,liu2015}, and consequently a small fraction of LISA-detectable MW disc DWDs may have eccentric orbits \citep{rajamuthu2025}. DWDs in GCs could also gain eccentricity through dynamical interactions in the dense stellar environments of GCs \citep{hills1975,heggie1975,heggie1996,willems2007}, and eccentricity has been proposed as a way of distinguishing the GW signals of DWDs in GCs from those in the field \citep{hellstrom2025}. Therefore, it would be insightful for future research to calculate LISA's sky location and distance measurement uncertainties for eccentric DWDs; however, \textsc{GWToolbox} cannot perform such calculations for eccentric binaries. This would be particularly relevant for GCs close to the Galactic Centre, whose binaries, as established in this paper, cannot be distinguished from MW disc DWDs from sky location and distance measurements alone, but may be distinguishable based on eccentricity.

\subsection{47 Tucanae}

Considering the sky locations and masses of the GCs in our sample, our results indicate that the most promising GC for finding a WD binary that could be resolved to that GC by LISA is 47 Tucanae. We note that 47 Tucanae is known, from EM observations, to host the object 47 Tuc X9 \citep{hertz1983,verbunt1998,grindlay2001}, a low-mass X-ray binary that is most likely a BH–WD binary, though it may alternatively be a NS–WD \citep{bahramian2017}. 47 Tuc X9 is the only known ultracompact BH–WD binary candidate in the MW; the masses of its components are uncertain, but its orbital period of c. 28 min \citep{bahramian2017} would place it within the LISA frequency band. The existence of this source supports our conclusion that 47 Tucanae is a promising GC for finding a LISA-resolvable WD binary.

\section{Conclusions}

We investigated whether LISA could resolve WD binaries in GCs around the MW to those GCs, based on the uncertainties in LISA's measurements of the sky location and distance of the binaries from their GW signals. To do this, we used the software package \textsc{GWToolbox} to simulate the sky location and distance measurement errors for test binaries at the locations of 20 of the most massive GCs around the MW. We compared these errors with those of binaries in the MW disc, using a MW model from the population synthesis code \textsc{SeBa}.

We find that LISA's ability to resolve the WD binaries to the GCs and distinguish them from the MW disc binaries strongly depends on the sky location of the GCs. Specifically, we can divide the GCs in our dataset into three classes: first, GCs where any detectable binary would be distinguishable from MW disc binaries with sky location measurements alone; second, GCs which have some overlapping MW disc binaries in terms of sky location but which should be distinguishable when taking into account distance measurements; and third, GCs which have so many overlapping MW disc binaries that it would not be possible to distinguish any individual binary in the GC from those in the MW disc, even with both sky location and distance measurements. GCs in the first class are located far away from the Galactic disc and those in the third class are close to the Galactic centre, while those in the second class have intermediate locations on the edge of the disc. Out of the 20 GCs in our dataset, we find that five are in the first category, five in the second and ten in the third. Overall, the most promising GC for finding a WD binary that could be resolved to that GC is 47 Tucanae, due to the combination of its sky location and large mass.

\section*{Data availability}

The GC catalogue from which we obtained the sky locations and distances of the GCs in our sample can be found at \url{ https://people.smp.uq.edu.au/HolgerBaumgardt/globular/}, and is summarised in Table \ref{param_table}. Other data underlying this article will be shared upon reasonable request to the corresponding author.

\begin{acknowledgements}

The authors thank Jan J. Eldridge for useful discussions. WGJvZ acknowledges support from Radboud University, the European Research Council (ERC) under the European Union’s Horizon 2020 research and innovation programme (grant agreement No.~725246), Dutch Research Council grant 639.043.514 and the University of Auckland.

\end{acknowledgements}

\bibliographystyle{aa}
\bibliography{references}

\begin{appendix}

\begin{table*}
\section{Summary parameters of our GC sample}
    \centering
    \caption{Summary parameters of our GC sample.}
    \begin{tabular}{c c | c c c}
    Index & Name & Right ascension (\degr) & Declination (\degr) & Distance (kpc) \\
    \hline
    (1) & $\omega$ Cen & 201.69699 & -47.47947 & 5.43 \\
    (2) & Terzan 5 & 267.02020 & -24.77906 & 6.62 \\
    (3) & Liller 1 & 263.35233 & -33.38956 & 8.06 \\
    (4) & 47 Tuc & 6.02379 & -72.08131 & 4.52 \\
    (5) & M19 & 255.65749 & -26.26797 & 8.34 \\
    (6) & M62 & 255.30415 & -30.11339 & 6.41 \\
    (7) & M14 & 264.40065 & -3.24592 & 9.14 \\
    (8) & M13 & 250.42181 & 36.45986 & 7.42 \\
    (9) & NGC 6440 & 267.22017 & -20.36042 & 8.25 \\
    (10) & M22 & 279.09976 & -23.90475 & 3.30 \\
    (11) & M5 & 229.63841 & 2.08103 & 7.48 \\
    (12) & M92 & 259.28076 & 43.13594 & 8.50 \\
    (13) & NGC 6380 & 263.61861 & -39.06953 & 9.61 \\
    (14) & M9 & 259.79909 & -18.51626 & 8.30 \\
    (15) & M28 & 276.13704 & -24.86985 & 5.37 \\
    (16) & NGC 6541 & 272.00983 & -43.71489 & 7.61 \\
    (17) & NGC 6553 & 272.32299 & -25.90807 & 5.33 \\
    (18) & NGC 362 & 15.80942 & -70.84878 & 8.33 \\
    (19) & NGC 6752 & 287.71710 & -59.98455 & 4.12 \\
    (20) & NGC 5927 & 232.00287 & -50.67303 & 8.27 \\
    
    \end{tabular}
    \tablefoot{Sky coordinates and distances of the 20 GCs in our sample, taken from the catalogues of \citet{catalogue_baumgardt1} and \citet{catalogue_baumgardt2} and references therein. Data retrieved from \url{ https://people.smp.uq.edu.au/HolgerBaumgardt/globular/} in January 2025. The indices are based on the ordering in Table C2 of \citet{wouter_gw_spectral}, from most massive to least massive; we note that some of the mass values in the Baumgardt online catalogue have been adjusted since \citet{wouter_gw_spectral} was published, but we used the same set of GCs for ease of comparison. Other parameters of the GCs can be found in Table C2 of \citet{wouter_gw_spectral}.}
    \label{param_table}
\end{table*}

\begin{figure*}
\section{Comparison of LISA measurement error calculations}
    \centering
    \includegraphics[width=\columnwidth]{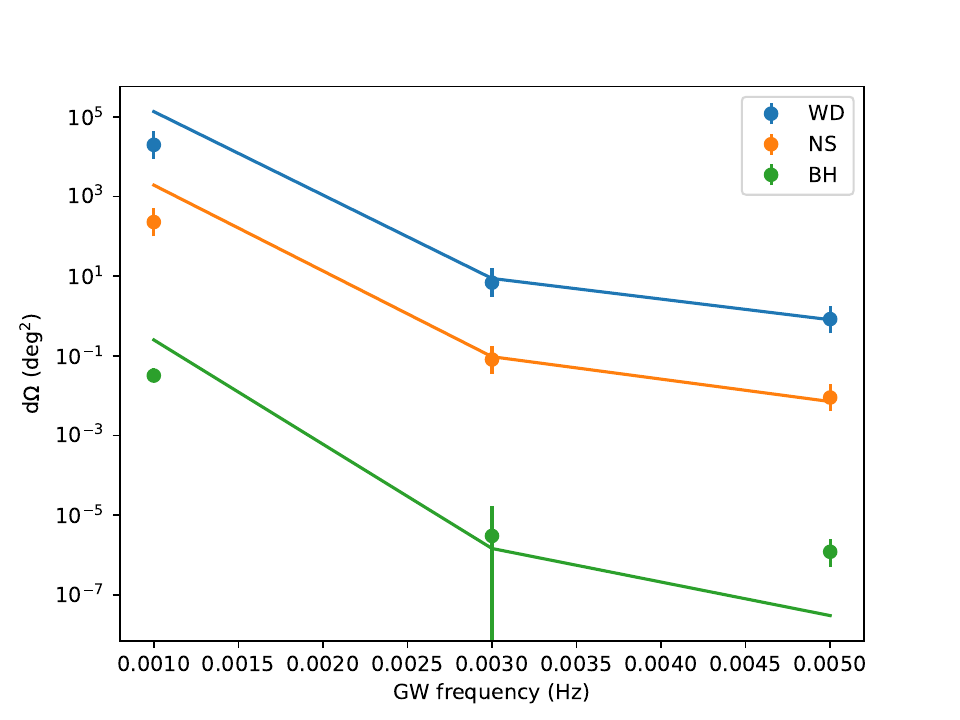}
    \includegraphics[width=\columnwidth]{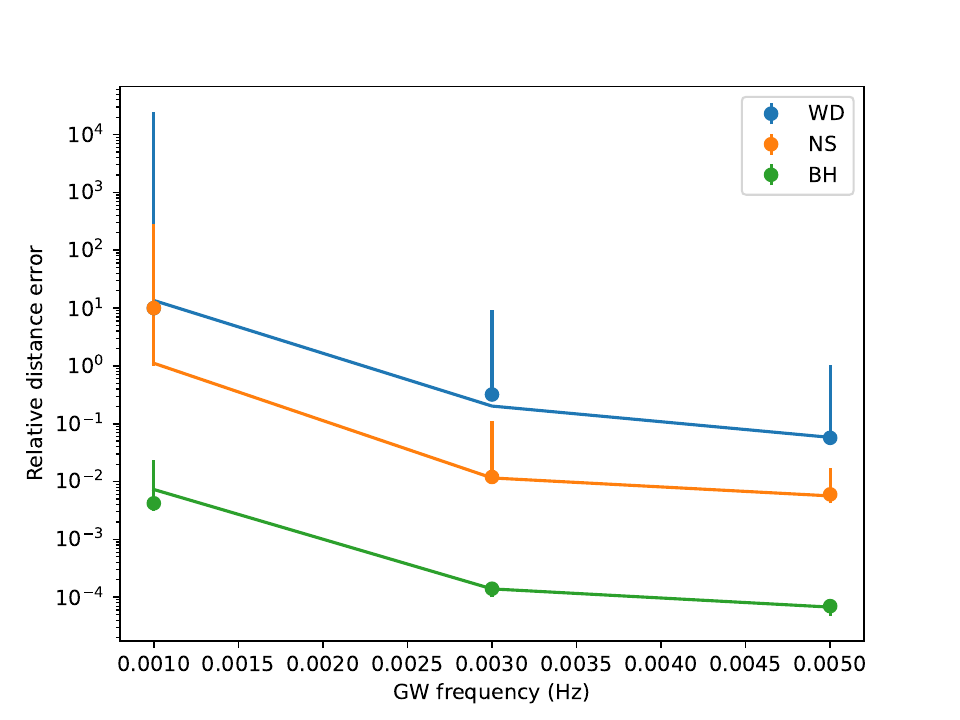}
    \caption{Comparing sky location and distance errors of GW-emitting binaries between \textsc{GWToolbox} and \citet{lisa_redbook}. The points show the data in Table 3.1 of \citet{lisa_redbook}: the sky location and distance errors of three example binaries. The lines show the errors for the same binaries as computed by \textsc{GWToolbox}, assuming an inclination of 0.6$\pi$ rad. The left plot shows the sky location error in deg$^2$, and the right plot shows the relative distance error (the distance error divided by the binary's distance itself). Full details on the example binaries can be found in Table 3.1 of \citet{lisa_redbook}.}
    \label{plot_redbook}
\end{figure*}

\end{appendix}

\end{document}